\documentclass[prd, 11pt,nofootinbib, superscriptaddress, preprintnumbers,floatfix]{revtex4}

\usepackage{amssymb,amsmath}
\usepackage{graphicx}
\usepackage{multirow}
\usepackage{comment}
\usepackage{color}
\begin{document}
\newcommand{\ignore}[1]{}
\def\mc#1{{\mathcal #1}}
\def\L{\Lambda}
\def\D{\Delta}
\def\d{\delta}
\def\a{\alpha}
\def\b{\beta}
\def\S{\Sigma}
\def\s{\sigma}
\def\e{\epsilon}
\def\O{\Omega}
\def\k{\kappa}
\def\g{\gamma}

\title{Tetraquark bound states in the heavy-light heavy-light system}

\author{Zachary S. Brown}
\email{zsbrown@email.wm.edu}
\affiliation{Department of Physics, College of William \& Mary, Williamsburg, VA 23187-8795}
\affiliation{Thomas Jefferson National Accelerator Facility, Newport News, VA 23606}

\author{Kostas Orginos}
\email{kostas@wm.edu}
\affiliation{Department of Physics, College of William \& Mary, Williamsburg, VA 23187-8795}
\affiliation{Thomas Jefferson National Accelerator Facility, Newport News, VA 23606}
\preprint{JLAB-THY-12-1643}

\begin{abstract}
A calculation of the interaction potential of two heavy-light mesons in lattice QCD is used to study the existence of tetraquark bound states.
The interaction potential of the tetraquark system is calculated on the lattice with 2+1 flavours of dynamical fermions with lattice interpolating fields constructed using \emph{colorwave} propagators. These  propagators provide a new method for constructing all-to-all  spatially smeared the interpolating fields, a technique which allows for a better overlap with the ground state wavefunction as well as reduced statistical noise.  Potentials are extracted for 24 distinct channels, and are fit with a phenomenological non-relativistic quark model potential, from which a determination of the existence of bound states is made via numerical solution of the two body radial Schr\"odinger equation.

\end{abstract}

\maketitle

%
%
\section{Introduction}
The calculation of hadronic forces from first principles allows insight into how interactions of the fundamental quark and gluonic degrees of freedom manifest themselves at the hadronic level. 
Lattice QCD is an excellent tool for calculating hadronic observables in the low energy regime. 
Although lattice calculations in euclidean space are not well suited for the study of real-time scattering processes, two methods can be used to extract interaction information from the lattice. 
One method, developed by L\"{u}scher \cite{Luscher1990}, relates the elastic scattering phase shift of a two particle system in a finite periodic box with the energy levels of the system. 
An alternate method, used in the present work, extracts the interaction energy as a function of hadron separation. This method is only applicable for systems of hadrons containing more than one heavy quarks which can be treated in the static approximation providing a definite spatial position for the hadrons.

Phenomenologically,  two heavy-light  meson systems (which we will denote as HLHL)  have become interesting in the study of tetraquark bound states \cite{Vijande2007a} \cite{Vijande2007b} \cite{Vijande2009}. 
It has long been known that the binding of a $\bar{Q} \bar{Q} q q $ (with $q=u,d$) system increases with the mass ratio of the heavy to light quark flavours \cite{Carlson1988}, thus $\bar{c} \bar{c} q q $ and $\bar{b} \bar{b} q q $ systems are excellent candidates in the search for exotic four quark bound states. 
In Ref. \cite{Vijande2009} a distinction was made between two types of tetraquark bound states: molecular, in which the four quarks exhibit a single physical two-meson (singlet-singlet) component, and the more exotic compact bound states. 
The latter would involve a complicated color space structure in which quark pairs form color vectors which then combine to form a colorless four quark state \cite{Vijande2009}.
In spite of this complicated color structure, compact bound states can be interpreted as a mixture of various two meson (color singlet) components \cite{Vijande2009sym}. 
The expected features that would characterize a molecular bound state would be a small binding energy and a bound state RMS radius greater than that of the sum of the two particle sizes, i.e.:
\begin{align}
\Delta_R \equiv \frac{RMS_{4q}}{RMS_{M_1} + RMS_{M_2}} > 1 \nonumber
\end{align}
A compact state, on the other hand, would be more tightly bound and have a smaller RMS radius than the molecular state. 
In Ref.~\cite{Ohkoda2012} doubly heavy four quark states were modeled as hadronic molecules interacting via a meson exchange potential. Several of the doubly bottom bound states were found to be deeply bound and spatially compact, making them excellent candidates for tetraquark bound states. 
It is with these ideas in mind that we may begin to search for the signature of compact bound states on the lattice. 

A recent lattice calculation of the HLHL interaction energy \cite{Wagner2011} has in fact hinted at the possibility of a bound tetraquark state in one channel that exhibits a particularly wide and deep potential well when compared with other channels, although no exhaustive determination of a bound state was undertaken. 
Our work goes beyond this presenting a quantitative determination of a bound state energy in the HLHL system from a lattice calculation. 

An inherent difficulty in making comparisons between theoretical models and lattice calculations performed in the static limit stems from the omission of the heavy quark spin in the static limit. 
As $m_H \rightarrow \infty$, the integer valued ($J=0,1$) angular momemtum eigenstates of a single heavy light meson map onto a single static limit eigenstate with $J=1/2$. 
The energies of the non-static angular momentum eigenstates also converge to a single energy corresponding to the $J=1/2$ eigenstate. 
Although the two spaces map onto each other, there is not a simple one to one correspondence between static limit eigenstates and their non-static counterparts, and care must be taken in making identifications between the two spaces. 
Previous lattice studies of HLHL interaction  energy (\cite{Michael1999}, \cite{Detmold2007} for example) performed in the quenched approximation and included uncontrolled systematic errors because of this.   
Recently dynamical quarks have been used to calculate the HLHL interaction energy using a complete set of quantum numbers which exploits the full set of symmetries of the HLHL system \cite{Wagner2010}. 

With our choice of quantum numbers (presented in section \ref{sec:Background}) we are able to draw a connection between the quantum numbers and the qualitative behavior of the states. 
Additionally, by way of symmetry arguments, we are able to relate our static-limit states to non-static angular momentum eigenstates. 

%
%
\section{Background}
\label{sec:Background}

\subsection{Heavy-Light states}
The quark model view of a heavy-light meson is of a heavy anti-quark $\bar{Q}$ coupled to a light quark $q$. 
The relevant quantum numbers to describe such a state are total angular momentum $J$ and its projection along some axis (here arbitrarily chosen to be $\hat{z}$) $J_z$, and the parity $P_i$ as well as the relevant flavor quantum numbers. 
For our interests, we choose $\bar{Q} = \bar{b}$ and $q = \{u,d\}$. 
Therefore all states then have bottomness $b=+1$, and are otherwise classified by total isospin and the third component of isospin $\left(I,I_z\right) = \left(1/2,\pm 1/2\right)$. 
Throughout this work, we make the assumption that we fit our correlation functions with a sufficiently large $t_{min}$ such that contributions from excited states have died out and we extract only the ground state energy.
Furthermore, we assume that states with non-zero orbital angular momentum $L$ are at sufficiently high energies as to have a negligible contribution to the ground state energies which we extract. 
We are then free to speak of the spin and total angular momentum interchangeably.

In heavy quark effective theory, spin dependent contributions enter into the heavy quark action at order $1/m_H$, and in the static limit ($m_H \rightarrow \infty$) the heavy quark acts as a static color source. 
This means that the spin of the HL meson comes only from the light degrees of freedom. 
Because of this, the physical HL meson states with $J = \left(0,1\right)$ become degenerate in the static limit, with both represented by a single $J=1/2$ state. 
The relevant angular momentum classification is then $\left(J,J_z\right) = \left(1/2,\pm1/2\right)$. 
With the above flavor assignments, the lowest energy excitations of the B spectrum with $J^P=\{0,1\}^-$ (coupling to the static $J^P=1/2^-$ B) are $B_{0,\pm}$ and $B^*$, and for $J^P = \{0,1\}^+$ (coupling to the static $J^P=1/2^+$ $B_1$), the ground state $B_1\left(5721\right)^0$ (neglecting excited states). 

\subsection{Heavy-Light Heavy-Light states}
When constructing states with a pair of HL mesons, care must be taken in determining a relevant set of quantum numbers that fully exploit the symmetries of the problem. 
The flavor quantum numbers for a Heavy-Light Heavy-Light (HLHL) system are straightforward, and for a $\bar{Q} q \bar{Q} q$ (with $q= \{u,d\}$) there are two isospin combinations, an isospin triplet with $I=1$ and an $I=0$ singlet.  
For a HLHL pair separated by a vector $\vec{r}$ the rotational symmetry is broken to rotations around the separation axis. 
Total angular momentum $J$ is therefore no longer a conserved quantity, though its projection along the axis of separation (arbitrarily taken to be $\hat{z}$) is still conserved. 
The system will also be symmetric or antisymmetric under parity  as well as reflections through a plane containing the separation axis, which we shall call $P_{\perp}$. 
This last transformation can be accomplished by a parity transformation followed by a rotation of $\pi$ about an axis perpendicular to the reflection plane. 
States with $J_z = \pm 1$ are not invariant under this transformation (being mapped onto each other), but their average is an eigenstate of $P_{\perp}$. 
Lastly we choose to classify HLHL states by intrinsic parity $P_i$, defined to be the product of the intrinsic parities of the two light quarks, and (full) parity $P$, defined as the product of the intrinsic parity transformation and coordinate inversion of the two particle spatial wavefunction.  
We will use both parity quantum numbers in our classification of states. 

%
%
\section{Methodology}

\subsection{HL and HLHL interpolating fields}

A general interpolating operator coupling to a single heavy-light state is given by:
\begin{equation}
\mathcal{O}_{HL}\left(\vec{x}\right)  = \bar{Q}\left(\vec{x}\right) \Gamma q\left(\vec{x}\right)
\end{equation}
with $\Gamma$ chosen to achieve the desired angular momentum and parity quantum numbers. 
For pseudoscalar HL states,  $\Gamma = \gamma_5, \gamma_i$ (with $i=1,2,3$), corresponding to a particle in the static limit with $J^P=1/2^-$, which we will refer to simply as $B$. 
$J = 1$ meson states with $\Gamma = 1, \gamma_i \gamma_5$ correspond to a state with $J^P = 1/2^+$, which we shall refer to as $B_1$.  
We make the choice $\Gamma = \gamma_5$ for $\mathcal{O}_B$ and $\Gamma = 1$ for  $\mathcal{O}_{B_1}$. 
As it will be useful in the analysis of HLHL states, it should be noted that for these choices of $\Gamma$, correlation functions constructed from $\mathcal{O}_B$ interpolating fields will consist of only upper (positive parity) components in the Dirac basis of the light quarks while those constructed from $\mathcal{O}_{B_1}$ will consist of only lower (negative parity) components. 
This is explicitly shown in Appendix \ref{App:Single_HL}. 
The states are classified by the additional flavor quantum numbers $\left(I,I_z\right) = \left(1,\pm1\right)$ for $q=\{u,d\}$.

For HLHL states, we want to create states with definite $\left(I,I_z,|J_z|, P_{\perp},  P,  P_i\right)$ and displacement $\vec{r}$ at the source and sink. To do this, we want to couple only our light quarks in spinor space to specify the quantum numbers of the state while allowing the heavy quarks to act only as color sources. Our general HLHL operator is then given by:
\begin{align}
 \mathcal{O}_{HLHL}^{\left(I,I_z,|J_z|, P_{\perp},  P,  P_i\right)}\left(\vec{x},\vec{r}\right) = \bar{Q}\left(\vec{x},t\right) \bar{Q}\left(\vec{x}+\vec{r},t\right) \times \left[q\left(\vec{x},t\right) q\left(\vec{x}+\vec{r},t\right)\right]\bigg|_{\left(I,I_z,|J_z|, P_{\perp},  P,  P_i\right)}
 \end{align}
 where the light quark wavefunctions $\left[q\left(\vec{x},t\right) q\left(\vec{x}+\vec{r},t\right)\right]$ are combined in such a way as to achieve the set of quantum numbers $\left(I,I_z,|J_z|, P_{\perp},  P,  P_i\right)$ of the system. 
The explicit construction of these wavefunctions is described in Appendix \ref{App:wfn}. 
For simplicity we restrict ourselves to identical source and sink interpolating fields neglecting any cross correlators between states. 
Isospin is a good quantum number on the $2+1$ flavor lattices with which we work, and we choose our interpolating fields to be isospin eigenstates with $\left(I,I_z\right) = \left(1,1\right)$ and $\left(I,I_z\right) = \left(0,0\right)$. 
At large spatial separations, we expect the energy of the four quark state to asymptotically approach the energy of it's dominant two meson component\footnote{Here we are referring to the dominant \emph{lowest energy} contribution, as we expect excited states to contribute negligibly to the extracted HLHL ground state energies}.
States with $P_i = -1$ will tend towards the energy of a $B B_1$ combination at large spatial separations.
There are two possible combinations of the light quark parities that yield $P_i=+1$: $\left(p_1,p_2\right)=\left(+,+\right),\left(-,-\right)$. 
In light of the above discussion of parity content of single HL states, we project our $P_i=+1$ interpolating fields to contain only negative or positive parity spinor components and retain these as distinct interpolating fields.
The expectation is that interpolating fields constructed from lower spinor components will exhibit a significantly higher ground state energy in relation to those constructed from upper components. The reason for this is that the $(-,-)$ interpolating field are constructed by the product of two $B_1$ meson interpolating fields, thus should exhibit an asymptotic energy (as $\vec{r} \rightarrow \infty$) near twice that of the single $B_1$ energy. Similarly the $(+,+)$ interpolating field is constructed from the product of two $B$ meson interpolating fields tending asymptotically as $\vec{r} \rightarrow \infty$ towards a ground state energy of twice that of a single $B$ meson.   
We differentiate all interpolating fields by their dominant asymptotic content in the tabulation of interpolating fields in Table \ref{tab:operators}.

\begin{table}[ht]
\centering

\begin{tabular}{|l|l|c|}
\hline
$\left( I,I_z\left| J_z\right|, P_{\perp}, P, P_{i}\right) $ & $\left( I,I_z,\left| J_z\right|, P_{\perp}, P, P_{i}\right) $ & Dominant asymptotic content \\
\hline
\hline
(1,1,1,--,--,+) &   (0,0,1,--,+,+) & $B B$ \\
 (1,1,0,--,--,+) &   (0,0,0,--,+,+) &  $B B$ \\
 (1,1,0,+,+,+) &   (0,0,0,+,--,+) &   $B B$ \\
 (1,1,1,--,--,+) &    (0,0,1,--,+,+) &  $B_1 B_1$ \\
 (1,1,0,--,--,+) &   (0,0,0,--,+,+) &  $B_1 B_1$ \\
 (1,1,0,+,+,+) &   (0,0,0,+,--,+) &  $B_1 B_1$ \\
 (1,1,1,+,+,--) &  (0,0,1,+,--,--) & $B B_1$ \\
 (1,1,0,+,+,--) &   (0,0,0,+,--,--) &  $B B_1$ \\
 (1,1,0,--,--,--) &  (0,0,0,--,+,--) &  $B B_1$ \\
 (1,1,1,+,--,--) & (0,0,1,+,+,--) & $B B_1$ \\
 (1,1,0,+,--,--) &   (0,0,0,+,+,--) &  $B B_1$ \\
 (1,1,0,--,+,--) &   (0,0,0,--,--,--) &  $B B_1$ \\ 
 
 \hline
\end{tabular}
\caption{HLHL interpolating operator basis and expected asymptotic values}
\label{tab:operators}
\end{table}

\section{Details of the lattice calculation}
We work with colorwave propagators (described below) calculated on $n_f = 2+1$ anisotropic ($24^3 \times 128 $) lattices generated by the Hadron Spectrum Collaboration \cite{Lattices} with a pion mass of roughly 380 MeV. 
The fermion action used was the clover Wilson action with stout link smearing, not smeared in the temporal direction. 
The gauge action was Symanzik tree level tadpole-improved without a rectangle in the temporal direction, preserving temporal ultra-locality. 
The spatial and temporal lattice spacings for these lattices are $a_s = 0.1227(8) $fm and $a_t = 0.03506(23) $fm. The pion mass on this ensemble is 0.0681(4) in temporal lattice units. 
The Chroma Software package for Lattice QCD \cite{Edwards} was used to generate both colorwave and heavy propagators. 
The calculation of the HL and HLHL energies was performed using 305 gauge field configurations with eight sources spaced evenly in the temporal direction.
Ground state energies were extracted using single exponential correlated fits, 
with an appropriate $t_{min}$ determined from  the quality of the  fit.

\subsection{Colorwave Formalism}

\subsubsection{Two quark states}
Consider a general operator for a two quark mesonic state:
\begin{equation}
\mathcal{O}\left(\vec{x}\right) = \bar{q}_1\left(\vec{x}\right) \Gamma q_2\left(\vec{x}\right) 
\end{equation}
where we assume for simplicity that the two quarks have different flavors.
We seek to calculate the correlation function with localized interpolating fields: (averaged over spatial source and sink locations to increase statistics)
\begin{eqnarray}
C\left(t,t_0\right) &=&\sum_{x,y}\left< \mathcal{O}\left(y\right) \mathcal{O}^{\dagger}\left(x\right)\right> \nonumber\\
                            &=& \sum_x \sum_y tr\left(	S_1\left(x,t_0|y,t\right) \Gamma S_2\left(y,t|x,t_0\right) \Gamma  	\right) 
\label{eq:cor_HL}
\end{eqnarray}
Following the methodology presented in \cite{Peardon2009}, we now consider any complete set of orthonormal states $\{ \phi_i\left(x\right) \} $ which satisfy:
\begin{equation}
\sum_i \phi_i^*\left(x\right)\phi_i\left(y\right) = \delta\left(x -y\right), \sum_x \phi_i^*\left(x\right)\phi_j\left(x\right) = \delta_{ij}  \;.
\label{eqn:com_rel}
\end{equation}
By inserting the completeness relation of eq.~\ref{eqn:com_rel} twice into the  two point function of eq.~\ref{eq:cor_HL}: 
\begin{align}
C \left(t,t_0\right) =& \sum_{x,x'} \sum_{y,y'} \left<S_1\left(x,t_0|y',t\right) \delta\left(y -y'\right)\Gamma S_2\left(y,t|x',_0\right) \delta\left(x -x'\right)\Gamma  	\right> \nonumber\\
								 =& \sum_{x,x'} \sum_{y,y'} \left<	S_1\left(x,t_0|y',t\right) \sum_i \phi^{*}_i\left(y\right)\phi_i\left(y'\right)\Gamma S_2\left(y,t|x',t_0\right)\sum_j \phi^*_j\left(x\right)\phi_j\left(x'\right)\Gamma  	\right> \nonumber\\
								 =& \sum_{i,j} S_1^{j,i} \left(t_0,t\right)\Gamma S_2^{i,j}\left(t,t_0\right) \Gamma  
\end{align}
where we have defined:
\begin{align}
 S^{i,j}\left(t,t_0\right) \equiv \sum_{x,y} \phi^*_i\left(y\right) S\left(y,t;x,t_0\right) \phi_j\left(x\right)
\label{eqn:basisprop}
\end{align}

\begin{figure}[H]
\centering
\includegraphics[height=6cm]{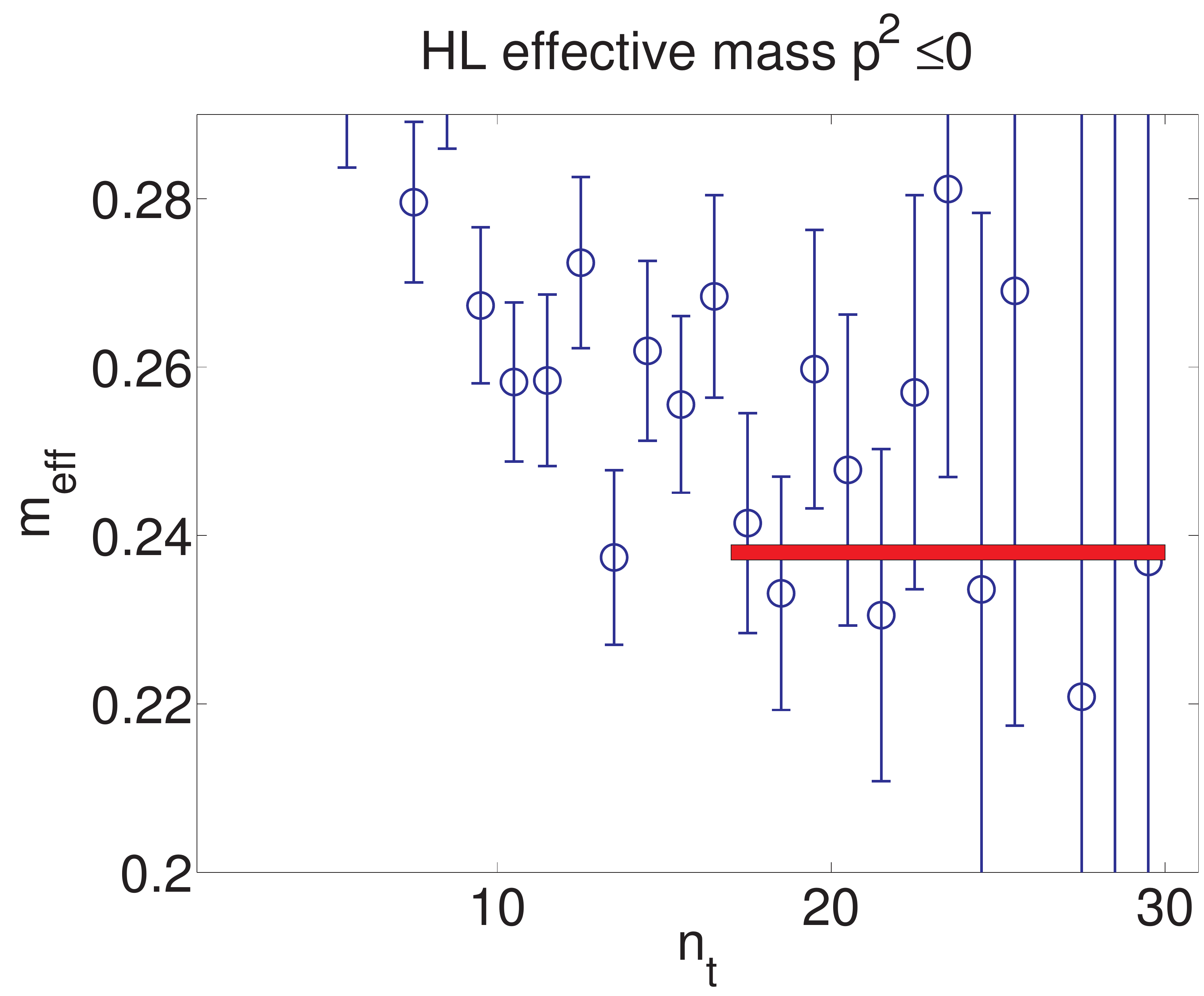}
\includegraphics[height=6cm]{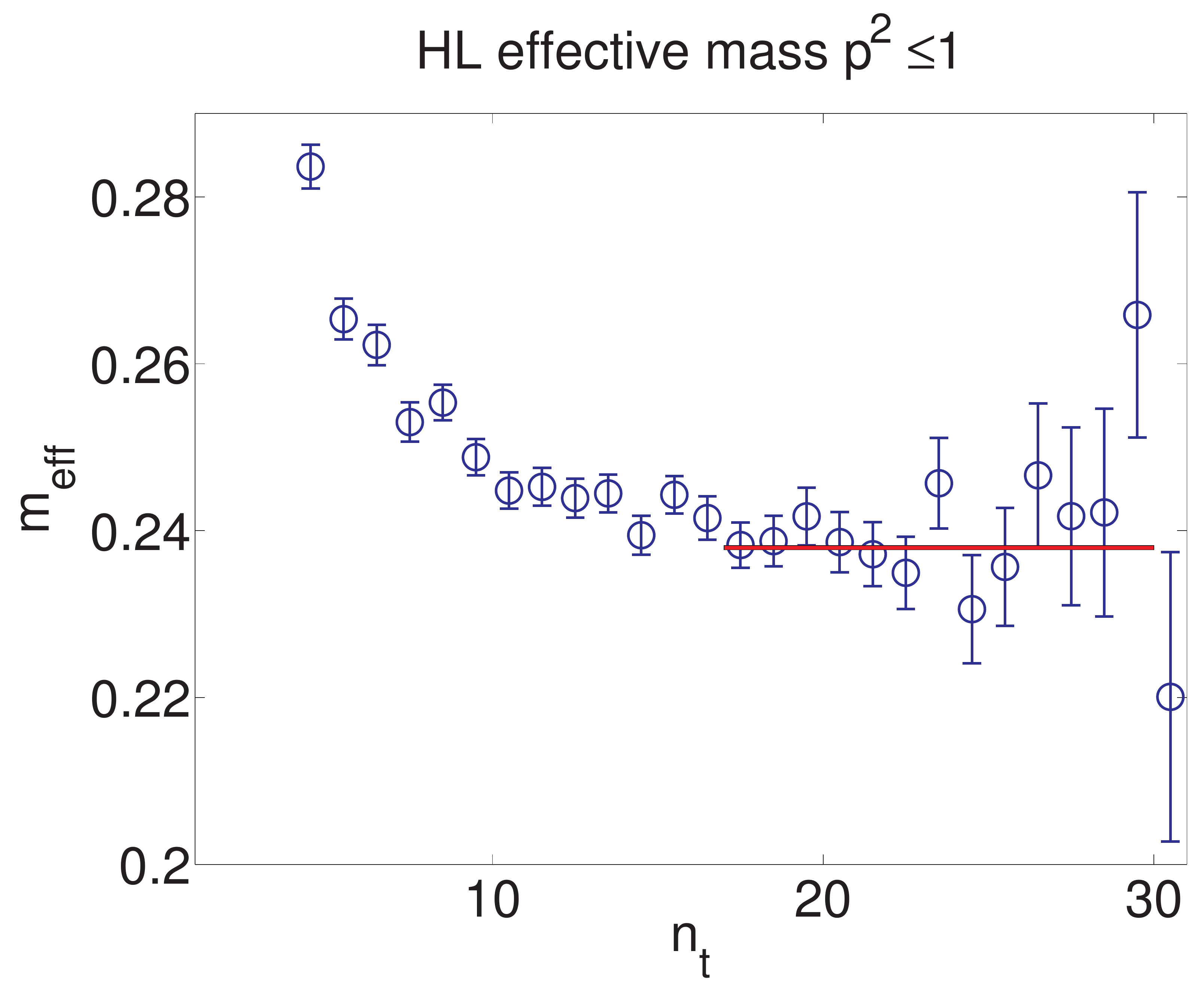} \\
\includegraphics[height=6cm]{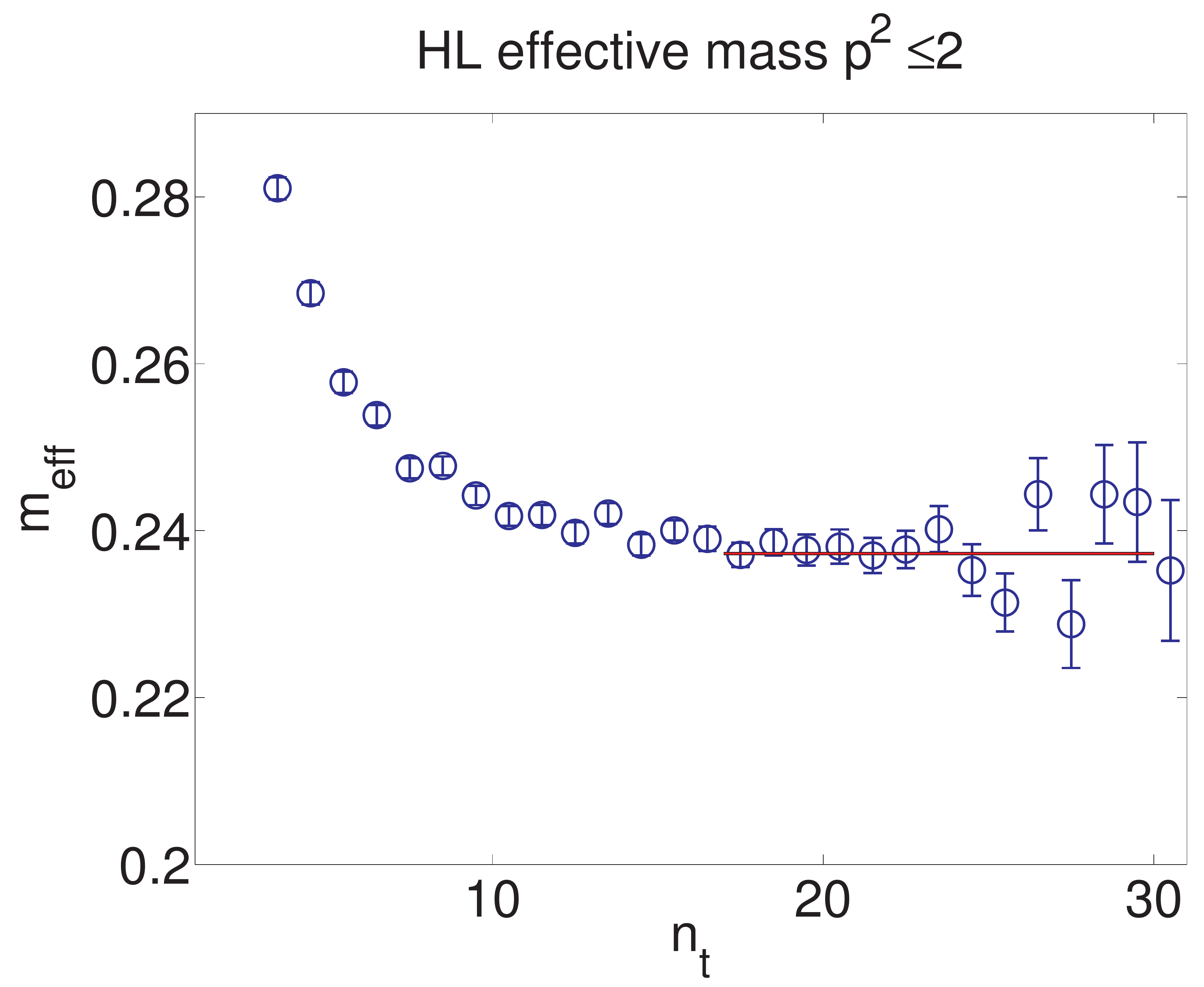}
\includegraphics[height=6cm]{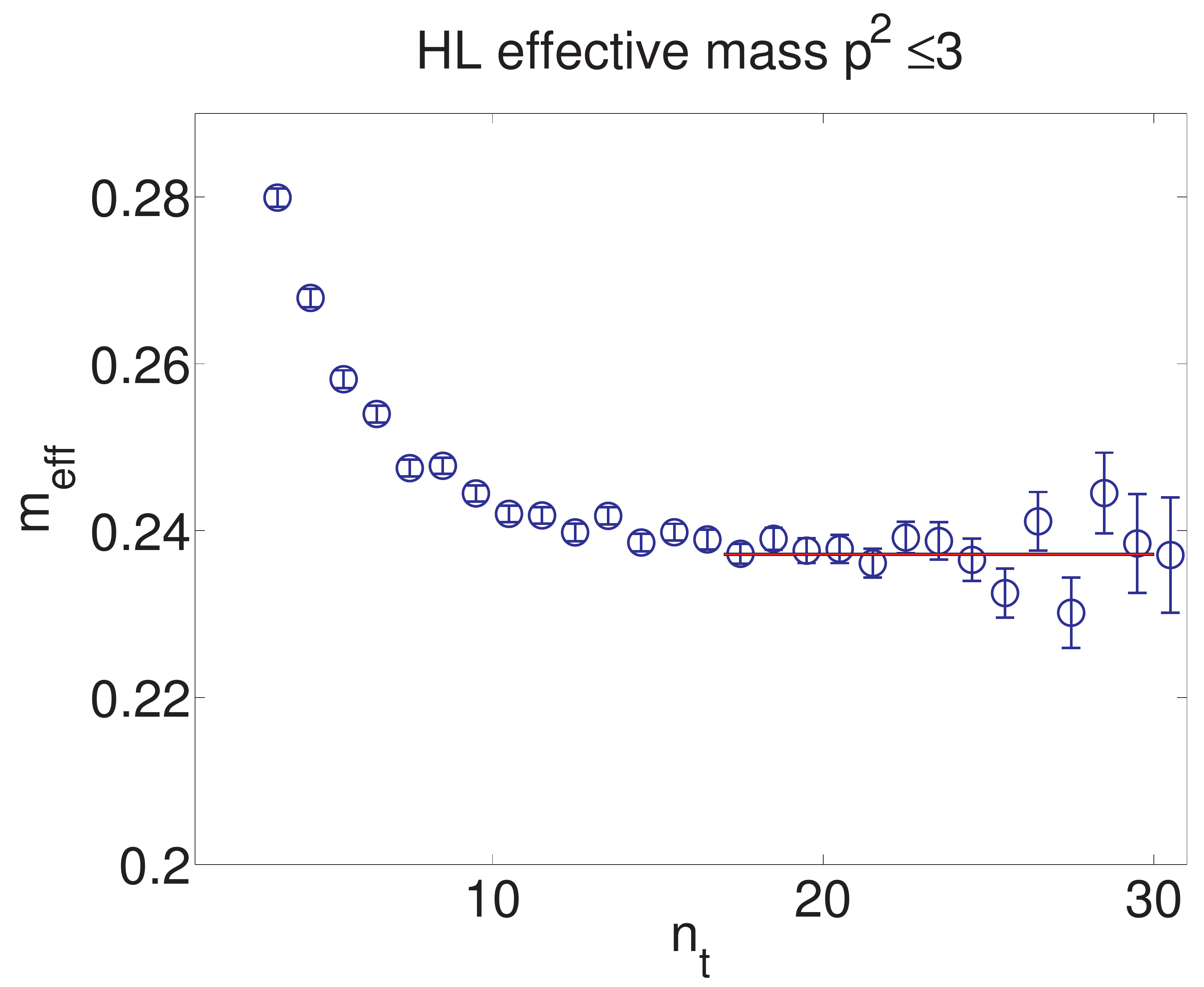} \\
\includegraphics[height=6cm]{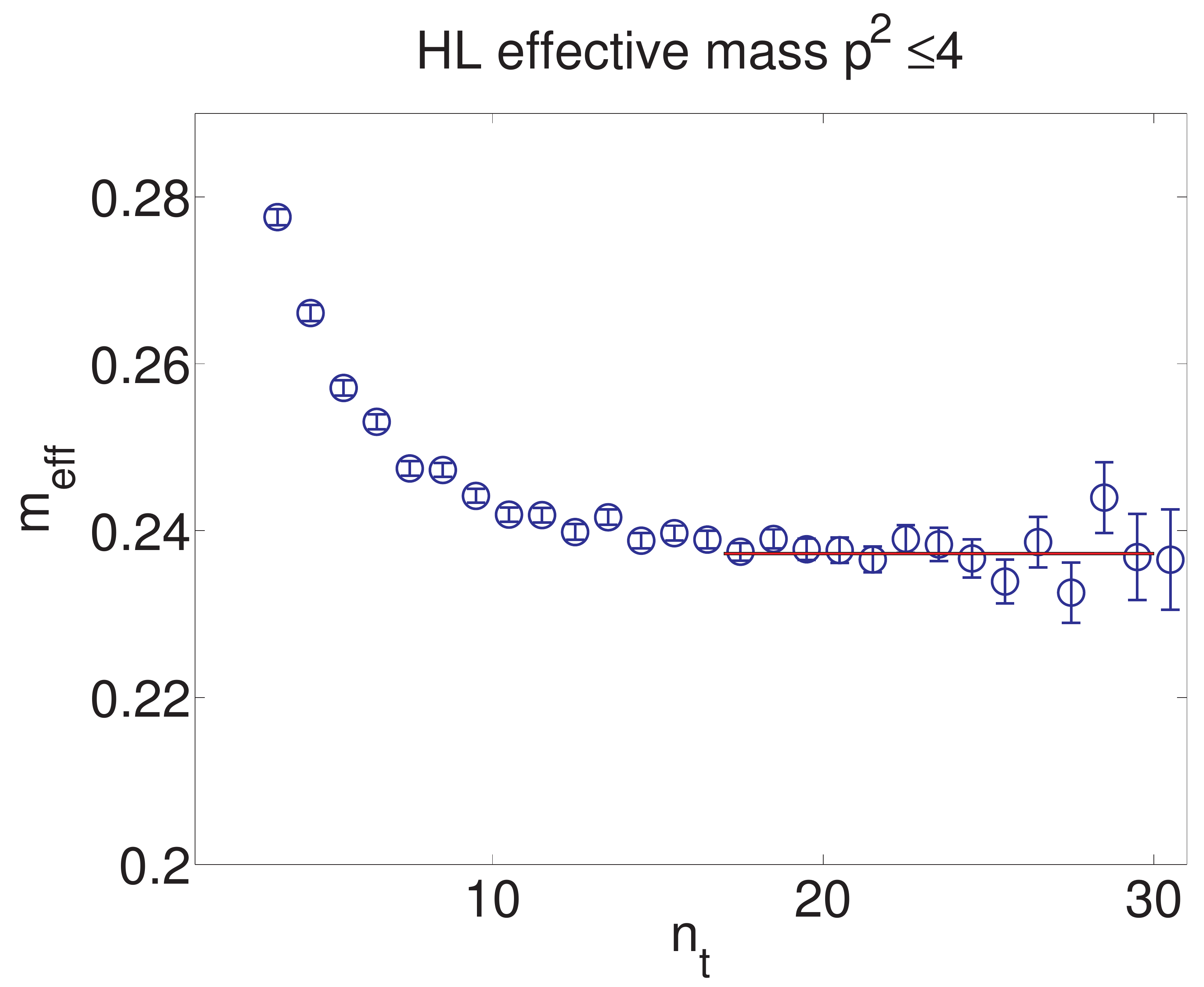}

\caption{Effective mass for HL $B$ for increasing $|p^2| \leq |p^2_{cut}|$} 
\label{fig:psqrcutlteq}
\end{figure}

A convenient choice for the $\{ \phi_i\left(x\right) \} $ is a plane wave basis: $\phi_i\left(x\right) \equiv \phi_p\left(x\right) = e^{-ipx}\delta_{s,s'}\delta_{c,c'}$. 
The delta functions here operate on color and spin.
With this choice of basis, we define $S^{i,j}\left(t,t_0\right) \equiv  S^{p,p'}\left(t,t_0\right)$ to be \emph{colorwave} propagators. 
The use of these propagators allows us to implement spatial smearing at the source and sink of our correlation functions.
In the limit where all momenta are summed over in equation \ref{eqn:basisprop}, all to all point-point propagators are recovered. 
However, introducing a maximum momentum cutoff $p^2_{cut}$ we are able to introduce and control the effective amount of spatial smearing\footnote{It should be noted that smearing  is achieved only in fixed gauge. In our case we use the Coulomb gauge, which is a smooth gauge allowing to project out high energy modes if the cutoff $p^2_{cut}$ is kept small .}. 
The effect of restricting the plane wave basis to $|p|^2 \leq p^2_{cut}$ (summing over a momentum space volume) is illustrated in Fig. \ref{fig:psqrcutlteq} where effective masses for single HL $B$ meson correlation functions\footnote{These HL correlation functions are defined in Appendix \ref{App:Single_HL}, eq.~\ref{eqn:HL_Corr}} are presented.  
It's evident that the noise of the signal decreases by increasing the momentum space cutoff (as this increases the statistics contributing to the correlation function).

Each effective mass plateau appears to begin at roughly the same point independent of $p^2_{cut}$, and thus a common fit range of $17 -30$ was chosen for all values of $p^2_{cut}$. 
In Fig. \ref{fig:psqrcutlteq} we can see that as $p^2_{cut}$ increases the overlap with excited states drops resulting lower values
for the effective mass at earlier times. This indicates that a small radial smearing of the quarks field results interpolating fields that have 
better overlap with the ground state of the system. 
Such behavior is likely due to the fact that the a non-relativistic HL meson in the static limit is a highly localized object whose wavefunction is confined to a small spatial region. 

In light of this behavior and in order to reduce computational cost associated with increasing the momentum cutoff, a value of $p^2_{cut} = 1$ was chosen for calculations of the HLHL system.  


\subsubsection{HLHL States}

We begin with a correlation function for two heavy-light mesons separated by $\vec{r}$ as described above:
\begin{align}
C_{HLHL} \left(t,\vec{r}\right)  =& \sum_x \left< \mathcal{O}_{HLHL}\left(\vec{x},\vec{r},t\right) \mathcal{O}^{\dagger}_{HLHL}\left(\vec{x},\vec{r},t_0\right) \right> \\
=& \sum_x \left<
\bar{Q}\left(\vec{x},t\right) \bar{Q}\left(\vec{x}+\vec{r},t\right) q\left(\vec{x},t\right) q\left(\vec{x}+\vec{r},t\right) 
\bar{q}\left(\vec{x}+\vec{r},t_0\right) \bar{q}\left(\vec{x},t_0\right)	Q\left(\vec{x}+\vec{r},t_0\right)	Q\left(\vec{x},t_0\right)									\right> \nonumber
\end{align}

Each heavy quark source can only be contracted with the sink at the same spatial location, and upon contraction we work only with the Wilson line portion of the heavy quark propagator, as we want the quantum numbers of the system to be determined entirely by the light degrees of freedom. There are two possible light quark contractions, one where the light quarks contract with source and sink at the same spatial location (direct), and one where the light quarks contract at the other spatial location (crossed). Performing these contractions, we have (omitting the overall color trace):
\begin{align}
C_{HLHL} \left(t,\vec{r}\right) =& \sum_x \gamma_5 W^\dagger\left(\vec{x};t,t_0\right)\gamma_5 \gamma_5 W^\dagger\left(\vec{x}+\vec{r};t,t_0\right)\gamma_5  \nonumber\\
											 &\times tr_d\left[ S\left(\vec{x}+\vec{r},t; \vec{x}+\vec{r},t_0\right) S\left(\vec{x},t; \vec{x},t_0\right) -			        		                                                   S\left(\vec{x}+\vec{r},t; \vec{x},t_0\right) S\left(\vec{x},t; \vec{x}+\vec{r},t_0\right)\right] 
\end{align}
Here, $tr_d$ denotes the trace over Dirac space spinor indices and $W$ is the Wilson line
\begin{equation}
W\left(\vec{x};t,t_0\right) = \prod_{t'=t_0}^{t} U^\dagger_4(\vec{x},t')
\label{eq:Wline}
\end{equation}

We now introduce our partially fourier transformed light quark propagators as:
\begin{align}
S\left( x'_1,t; x_1,t_0 \right) = \sum_{p'_1, p_1} e^{i p'_1 x'_1} S\left( p'_1,t; p_1,t_0 \right)e^{-i p_1 x_1}
\end{align}
where sums over momenta $p_i$ have been restricted to $|p^2| \leq 1$ as described in the previous section. 

Using this, the above correlator can be rewritten as:
\begin{align}
C_{HLHL} \left(t,\vec{r}\right) = \sum_{p_1 p_1' p_2 p_2'}\sum_x &  \; \gamma_5 W^\dagger\left(\vec{x};t,t_0\right) \gamma_5 \gamma_5 W^\dagger\left(\vec{x}+\vec{r};t,t_0\right)\gamma_5 \times
												e^{i (p'_1 - p_1 + p'_2 - p_2) x}  e^{i(p'_2 - p_2) r}	 \nonumber\\
											 &\times \left[ S\left(p_2',t; p_2,t_0\right) S\left(p_1',t; p_1,t_0\right) -			
											 S\left(p_2',t; p_1,t_0\right) S\left(p_1',t;p_2,t_0\right)\right]  
\end{align}
Defining 
\begin{align} 
\mathcal{D}\left(\vec{r},t,t_0,\omega\right) \equiv \sum_x \gamma_5 W^\dagger\left(\vec{x};t,t_0\right)\gamma_5 \gamma_5 W^\dagger\left(\vec{x}+\vec{r};t_0,t\right)\gamma_5 e^{i (\omega) x} 
\end{align}
 with $\omega \equiv p'_1 - p_1 + p'_2 - p_2  $, our the final form of our HLHL correlation function becomes:
\begin{align}											 
C_{HLHL} \left(t,\vec{r}\right) = \sum_{p_1 p_1' p_2 p_2'} & \mathcal{D}\left(\vec{r},t,t_0,\omega\right) \times
												 e^{i(p'_2 - p_2) r} \nonumber\\
											 &\times \left[ S\left(p_2',t; p_2,t_0\right) S\left(p_1',t; p_1,t_0\right) -			        		                          
											                       S\left(p_2',t; p_1,t_0\right) S\left(p_1',t; p_2,t_0\right)\right ]  
\end{align}

With this method, we calculate the costly $ \mathcal{D}\left(\vec{r},t,t_0,\omega\right) $ first using a parallel code (parallelization over space time)  and then perform the far less expensive contractions with the colorwave propagators for our complete operator basis
on a scalar workstation class machine. 

%
%

\section{HLHL results}

For $q=\{u,d\}$ we have 24 unique HLHL corresponding to the operators enumerated in Table \ref{tab:operators}. Each potential curve is calculated by taking the jackknife difference between the energy of the HLHL state for various $\vec{r}$ and the energy of the expected two meson asymptotic state:
\begin{align}
V\left(\vec{r}\right) = E_{HLHL}\left(\vec{r}\right) - E_{B_{(1)}} - E_{B_{(1)}}
\end{align}
The statistical uncertainty for each point is determined from jackknife statistical analysis. The systematic uncertainties are determined by adjusting the chosen fit range by one time slice in each direction and averaging the observed deviations in the energy. The systematic uncertainty for both $E_{HLHL}$ and $E_{B_{(1)}}$ are determined independently and then added in quadrature to determine the systematic uncertainty on $V\left(r\right)$.

\begin{figure}[]
\centering
\includegraphics[height=6cm]{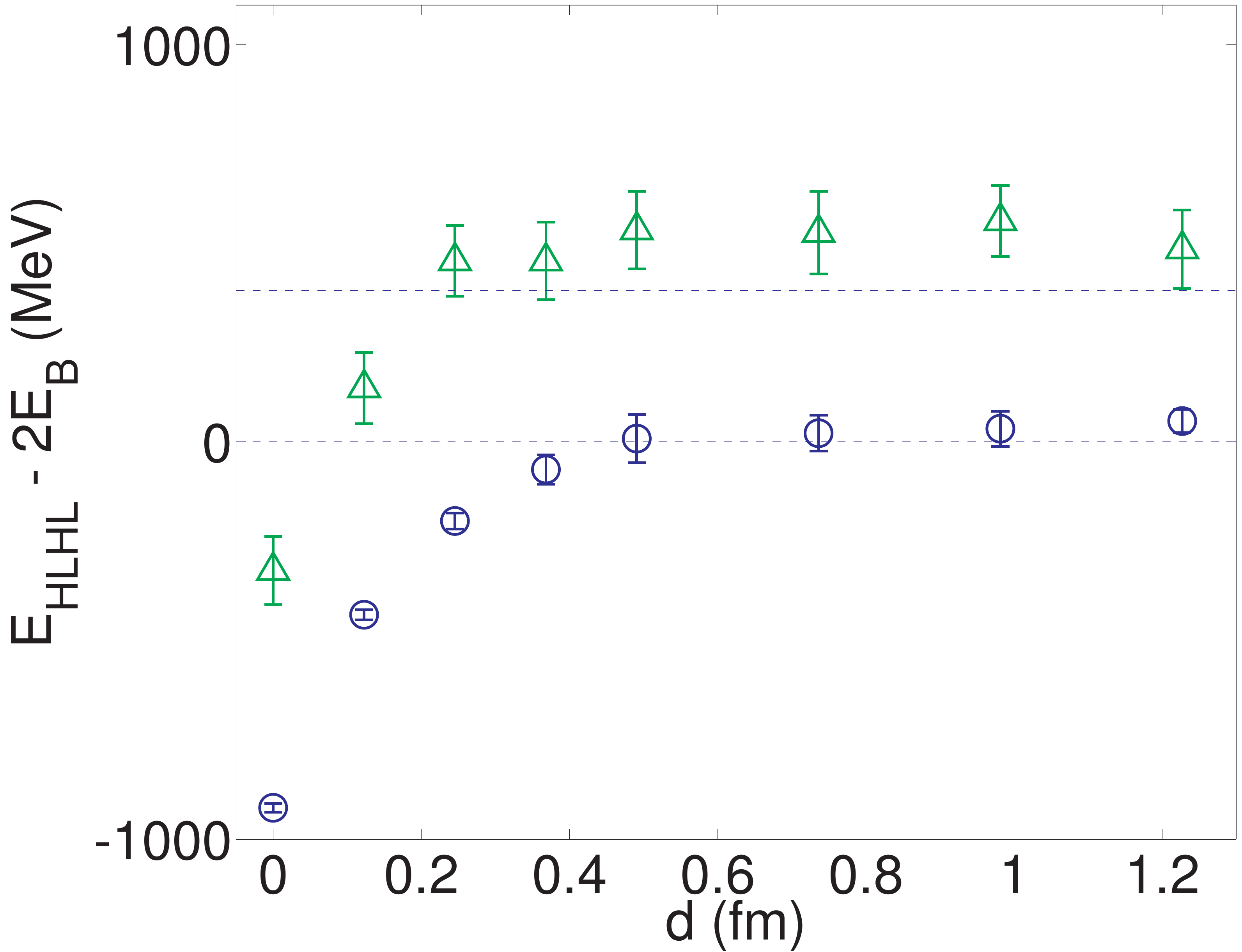}
\includegraphics[height=6cm]{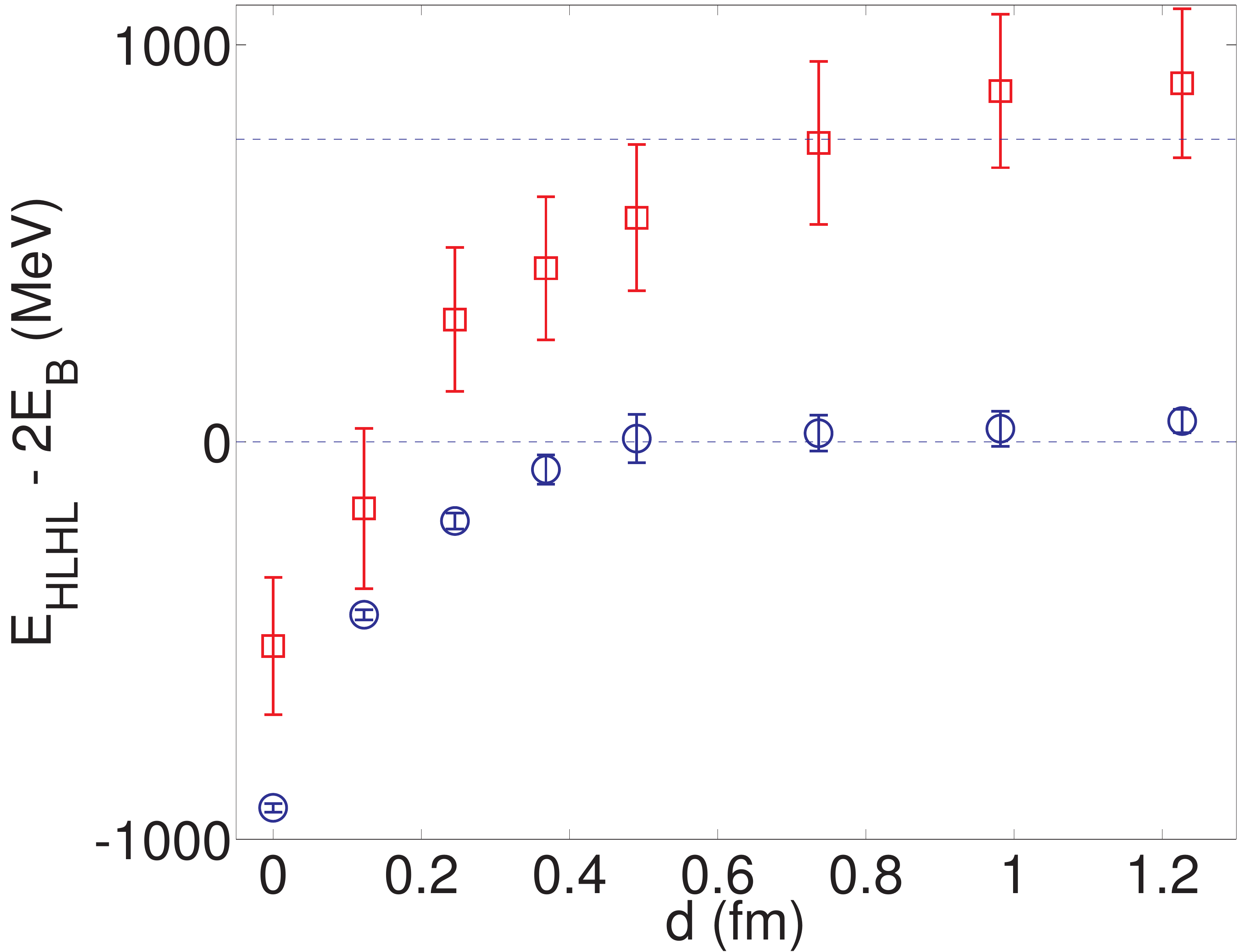}
\caption{Comparison of $B B$ vs. $B B_1$ (left) and $B B$ vs $B_1 B_1$ (right) asymptotic states. Here we take the energy difference for the three potential curves with respect to twice the HL $B$ energy}
\label{fig:AsymVals}
\end{figure}

We find three different asymptotic values for the various states as illustrated in Fig. [\ref{fig:AsymVals}]. The lowest lying asymptotic value corresponds to states with a positive intrinsic parity $P_i$ with all spin components in the correlation function projected to the upper spin components, while the highest asymptotic value corresponds to states with positive intrinsic parity and all spins projected to the lower components. This asymptotic behavior is in line with our expectation that the spin projection of our positive intrinsic parity operators helps to increase the coupling to the lower energy $B B$ state or the higher energy $B_1 B_1$ state.
The energy difference between the highest and lowest asymptotic values is roughly twice the energy difference between the single HL $B$ and $B_1$ states, indicating that they are both tending asymptotically towards their expected two meson asymptotic energies at long distances. 
The slight overshoot of the highest asymptotic state beyond it's expected value of twice the $B_1$ energy for $d > 0.8$ fm 
may be indicative of contamination from mixing of the HL $B_1$ with a $\pi - B$ state. All $P_i = (-)$ states exhibit an asymptotic tendency towards the sum of the single HL $B$ and $B_1$ energies as expected.

\begin{figure}[]
\centering
\includegraphics[height=6cm]{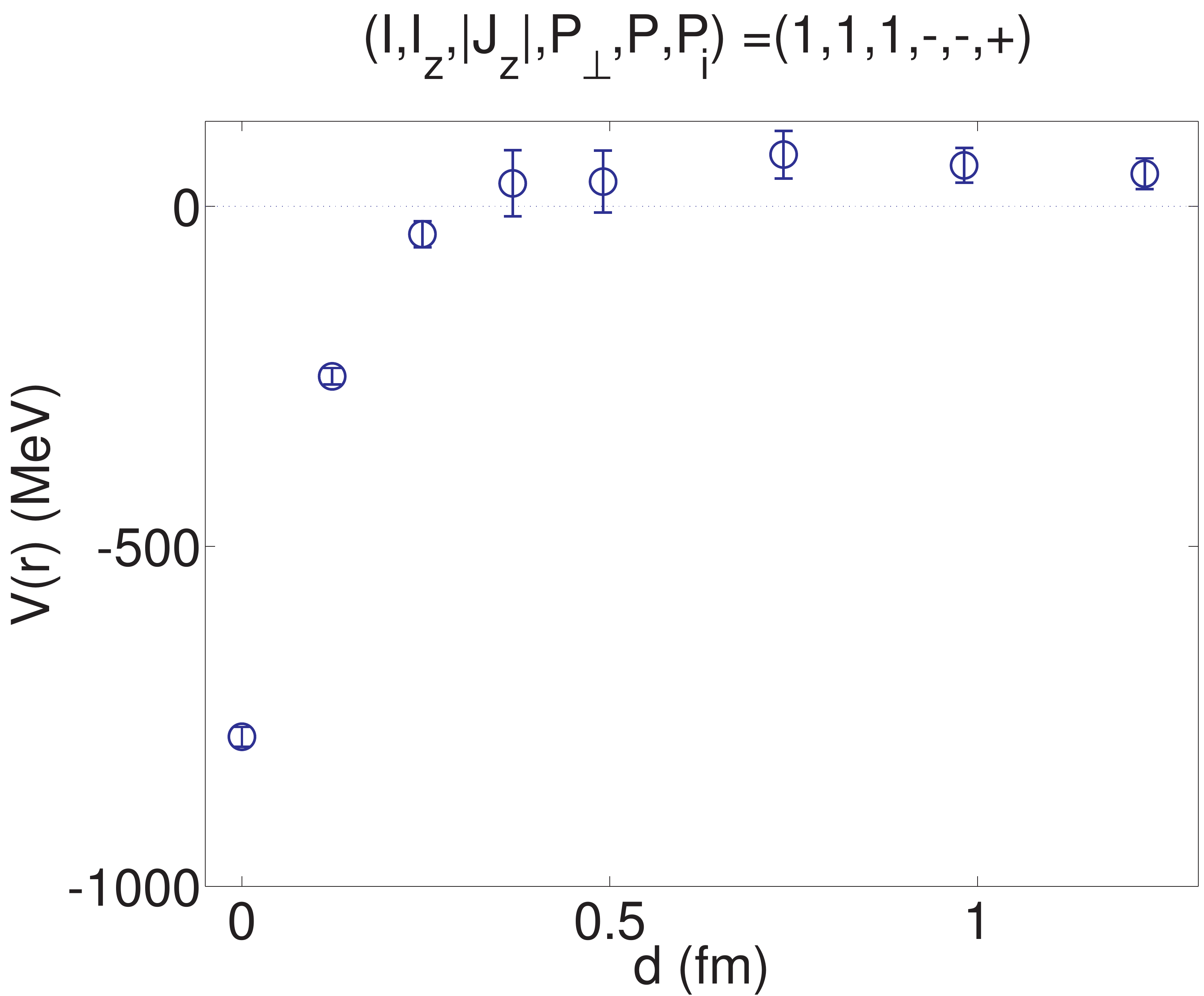}
\includegraphics[height=6cm]{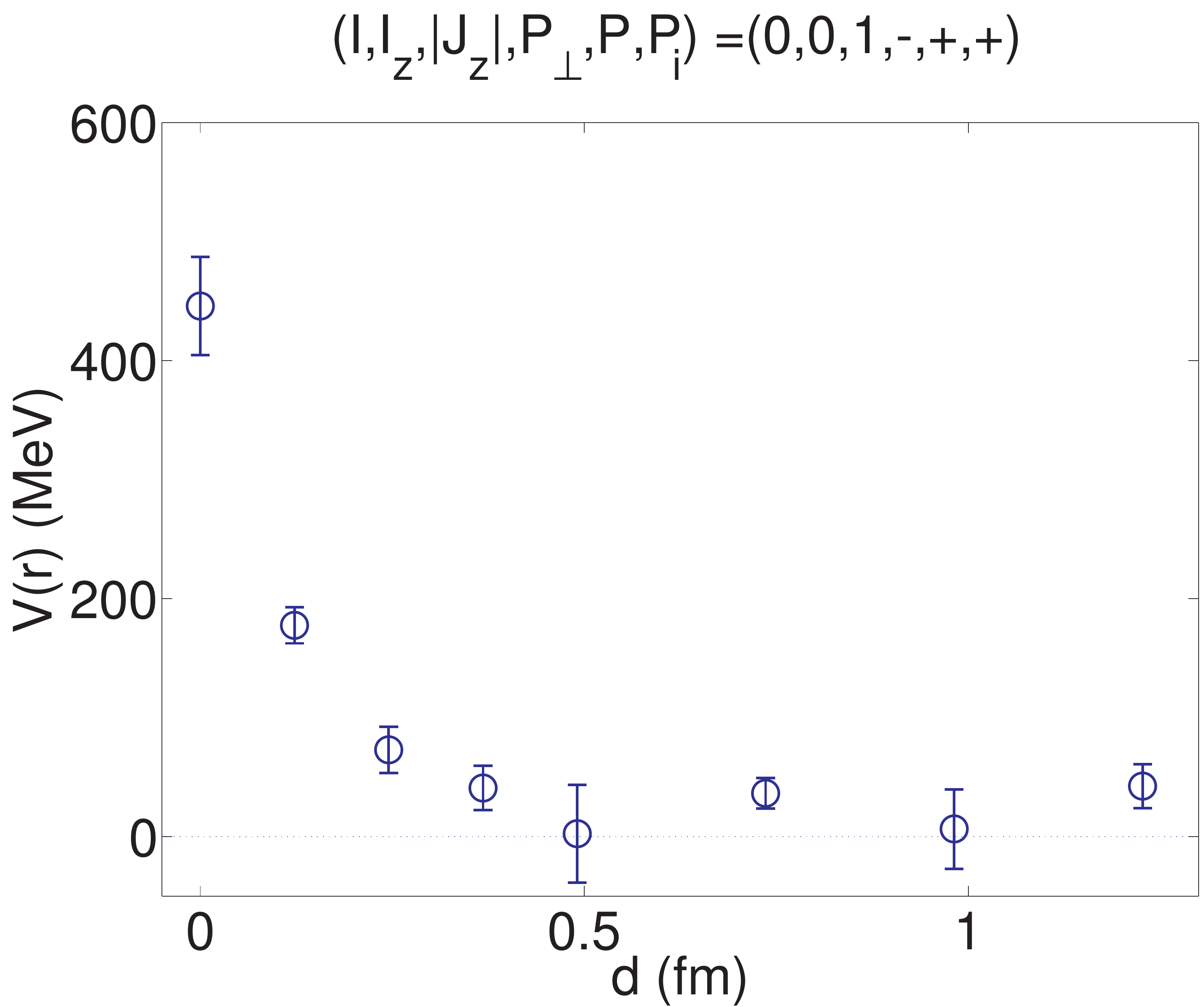} 
\vspace*{0.25in}\\
\includegraphics[height=6cm]{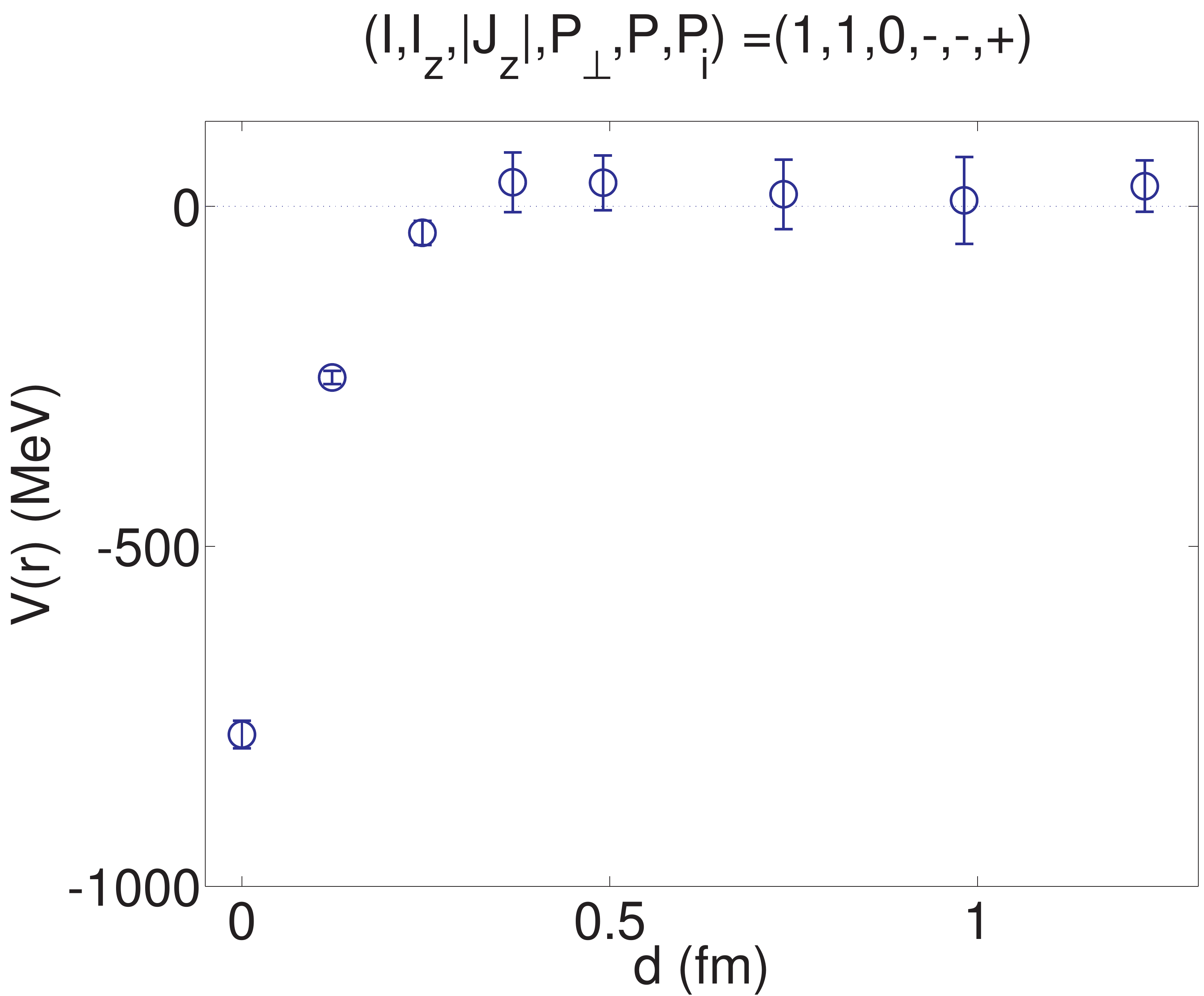}
\includegraphics[height=6cm]{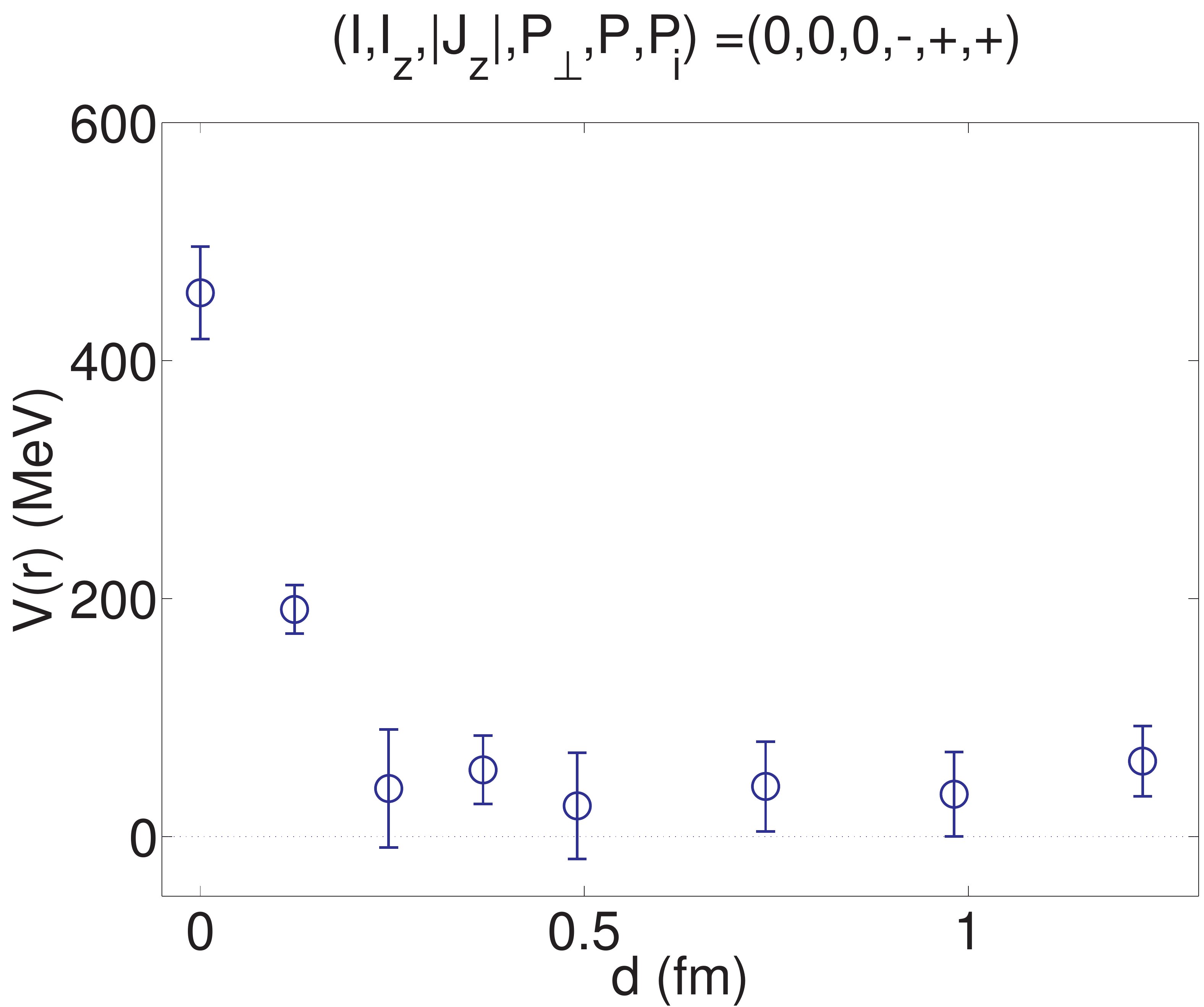}
\vspace*{0.25in}\\
\includegraphics[height=6cm]{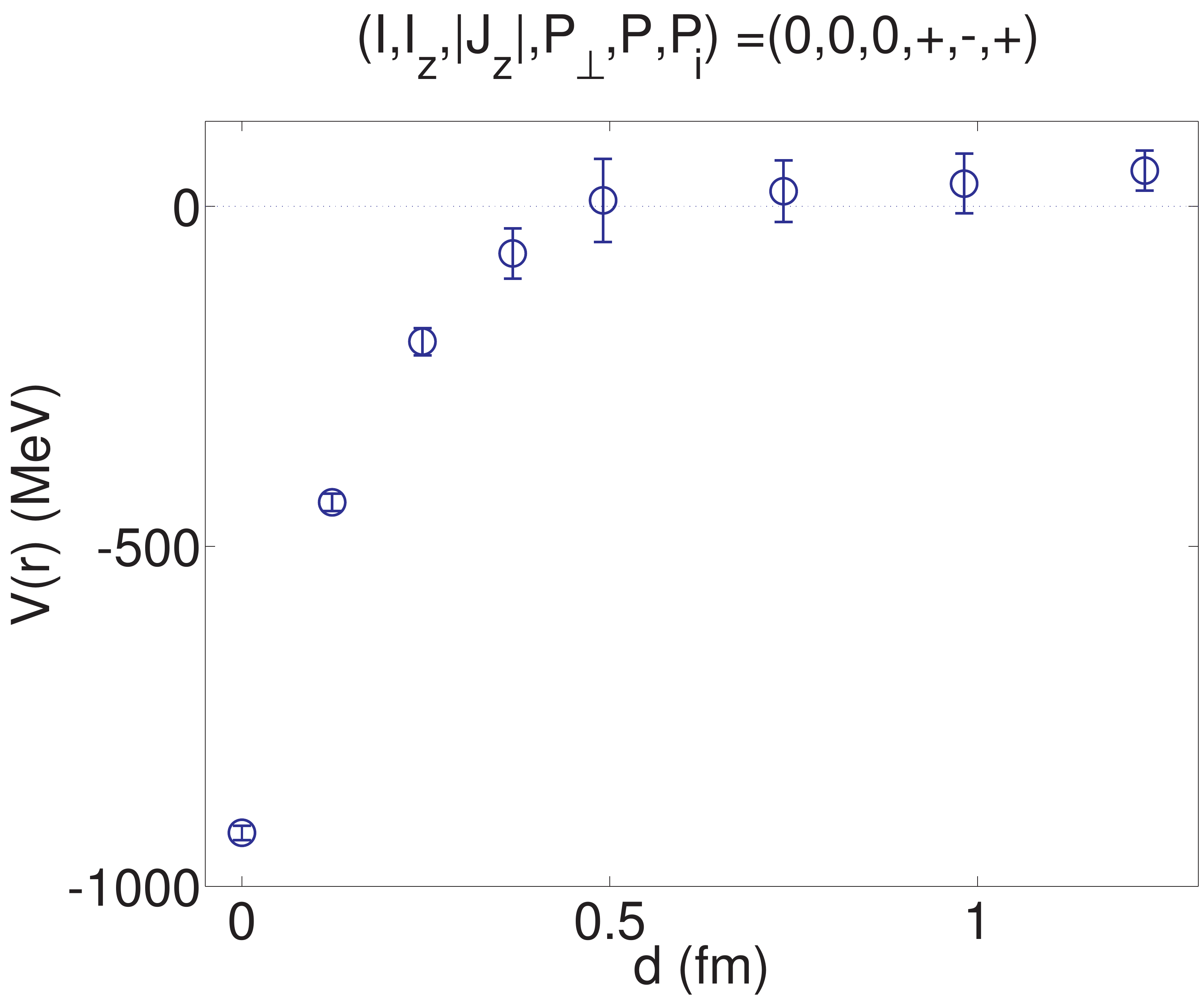}
\includegraphics[height=6cm]{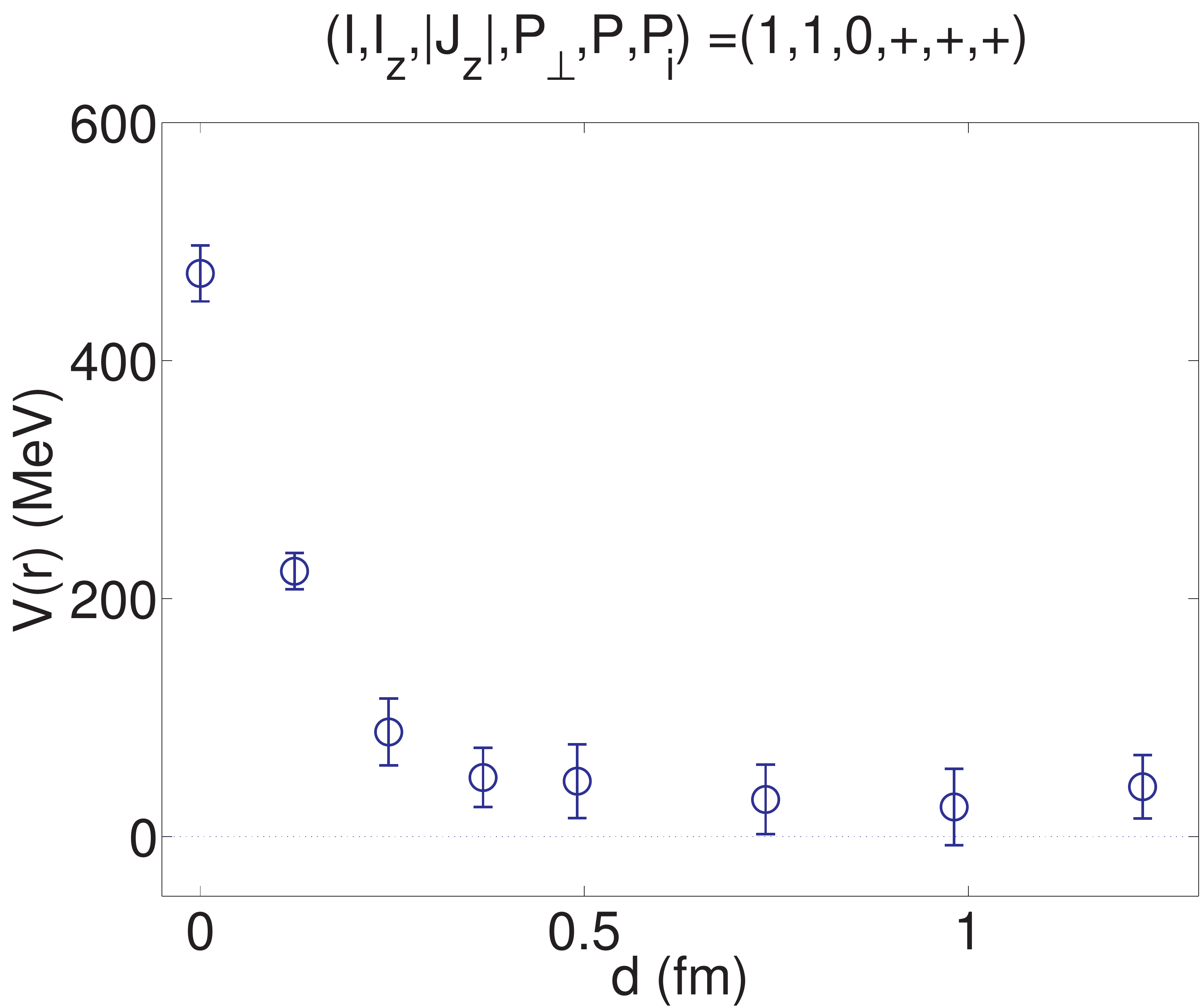}\\
\caption{Calculated HLHL $B B$ energies with expected asymptotic value (twice the calculated HL $B$ mass)}
\label{fig:BB_IntEn}
\end{figure}

As the states with the lowest asymptotic energy values trend most cleanly towards their expected asymptotic value (indicating the least contamination from excited states), we will focus mainly on these states which we present in Fig. [\ref{fig:BB_IntEn}]. 


Several aspects of these potential curves should be noted: First, we find that the product of exchange parity $P$ and intrinsic parity $P_i$, which is the symmetry of the two meson spatial wavefunction under spatial inversion, directly corresponds to the attractiveness $(-)$ or repulsiveness $(+)$ of the state. This is in agreement with \cite{Wagner2011}. Second, the $\left(I,I_z,|J_z|,P_{\perp},P,P_i\right) = \left(0,0,0\right) + - +$ exhibits a significantly deeper and wider potential well when compared with the two other attractive channels. This qualitative difference was acknowledged in \cite{Wagner2011}, and the quantum numbers of this channel are consistent with a bound state predicted in a phenomenological model in \cite{Vijande2009}.

\section{Bound States}
As the HLHL system has been predicted to be an excellent candidate for bound tetraquark states, we seek a quantitative method for extracting such a bound state (if one exists) from our lattice calculation. Our method is as follows: We fit our lattice potential to a phenomenological quark model potential as described in \cite{Barnes}. We make the choice to focus on the $\left(I,I_z,|J_z|,P_{\perp},P,P_i\right) = \left(0,0,0,+,-,+\right)$ channel, as previous work 
has hinted at the possibility of a bound state here. As a control, we also perform the fit for the $\left(I,I_z,|J_z|,P_{\perp},P,P_i\right)= \left(1,1,0,-,-,+\right)$ channel as well. 
In our fit, we neglect the $\vec{r} = 0$ points as the finite value of the potential at $\vec{r} = 0$ is a lattice artifact stemming from the ultraviolet cut off introduced by the lattice discretization, leaving us with 7 data points for each potential curve, and two free parameters from the fit model.
The model with the extracted fit parameters is then taken to be the interaction potential between two B mesons in the continuum limit. The two body (one-dimensional) Schrodinger equation is then solved numerically with this interaction potential to determine the existence of any negative energy (bound) states. It should be noted here that the solutions to the Schrodinger equation will converge to their continuum values as the continuum limit of the lattice calculation is taken. As we have only a single lattice spacing available to work with this continuum extrapolation is not an option, and it should be understood that the results presented in this section are at finite lattice spacing. 

\subsection{Potential Model}

We have limited our displacements $|\vec{r}| \leq 1.27$ fm, therefore long range effective interactions due to meson exchange do not provide a good description of the HLHL system. In reference \cite{Barnes}, a quark model picture of a two meson interaction was used to derive an interaction potential for the HLHL system, which included color coulomb, spin-spin, linear confinement interactions. Details of the derivation of the potential model can be found in the aforementioned reference, and we will only highlight several modifications we make when fitting this potential model to our numerical results. The quark model HLHL potential has the form:

\begin{align}
V_{BBDS}\left(r\right) = C_I V_{cc}\left(\alpha_s,\beta,r\right) + C_{\bold{S} \cdot \bold{S}} V_{ss}\left(\alpha_s,\beta,\bar{m},r\right) + C_I V_{lc}\left(b,\beta,r\right) 
\label{eqn:pot_mod}
\end{align}
with:
\begin{align}
V_{cc}\left(\alpha_s,\beta,r\right) &= \frac{-4\alpha_s}{9r} \left[1+\left(\frac{2}{\pi}\right)^{1/2} \beta r - 4\text{Erf}\left(\frac{\beta r}{2}\right)\right]e^{-\beta^2 r^2/2} \\ 
V_{ss}\left(\alpha_s,\beta,\bar{m},r\right) &= \frac{2}{27} \left(\frac{2}{\pi}\right)^{1/2} \frac{\alpha_s \beta^3}{\bar{m}^2}e^{-\beta^2 r^2/2} \\
V_{lc}\left(b,\beta,r\right) &= \frac{b}{3 \beta} \bigg{[} \beta r e^{-\beta^2 r^2/2} + 2\left(\frac{2}{\pi}\right)^{1/2} e^{-\beta^2 r^2/2} \nonumber  \\
\:\:\:\:\:\: \:\:\:\:\: &-\left(\beta r + \frac{2}{\beta r}\right) \text{Erf}\left(\frac{\beta r}{2}\right)e^{-\beta^2 r^2/2} - \frac{2}{\pi^{1/2}}e^{-3\beta^2 r^2/4}\bigg{]}  
\end{align}

Here, $\alpha_s$ is the strong coupling constant, $\beta$ is the spatial width of the quark model single HL meson wavefunction, $\bar{m}$ is the mass of the light quark in the $\overline{MS}$ scheme, and $b$ is the QCD string tension. The coefficients $C_I$ and $C_{\bold{S} \cdot \bold{S}}$, which contain the spin information of the HLHL state, are defined as matrix elements between initial (unprimed) and final (primed) two meson states and will be discussed further below.
It should be noted that the above potential model acquires an overall minus sign if the isospin wavefunction of the two meson state is antisymmetric. Additionally, the potential is a function of $|\vec{r}|$ and not $\vec{r}$, as any tensor interaction terms are neglected in this model. 

\subsection{Fit Model}
When applying the above model to our lattice data, we must make several modifications to the above quark model potential. 
Due to the use of periodic boundary conditions in the calculation, interactions with image ``charges''  lying past the boundary must be accounted for. 
We must also consider the possibility that there will be long range meson exchange interactions that were neglected in our choice of potential model. To account for these long range interactions, we extend the original model by adding a simple Yukawa like term for one pion exchange:
\begin{equation}
V^{Yuk}\left(r\right)  = V_{BBDS}\left(r\right) + g\frac{e^{-m_\pi r}}{r} 
\label{eq:ourModel}
\end{equation}
Here we take $m_{\pi}$ to be the mass of the pion on the gauge field configurations used in the calculation ($\sim 390$ GeV). The parameter $g$ is discussed below.

\begin{figure}[]
\centering
\includegraphics[height=6cm]{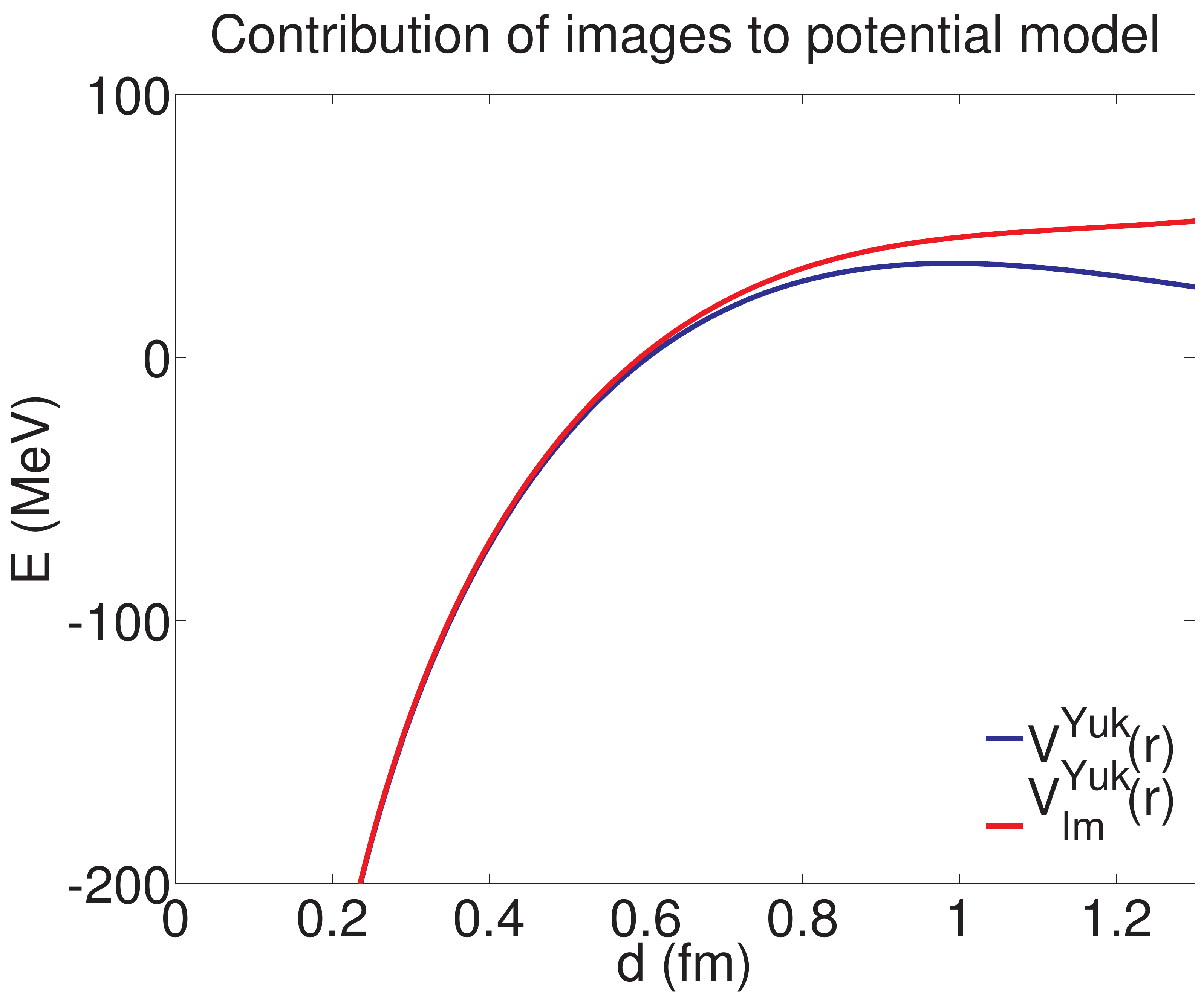}
\includegraphics[height=6cm]{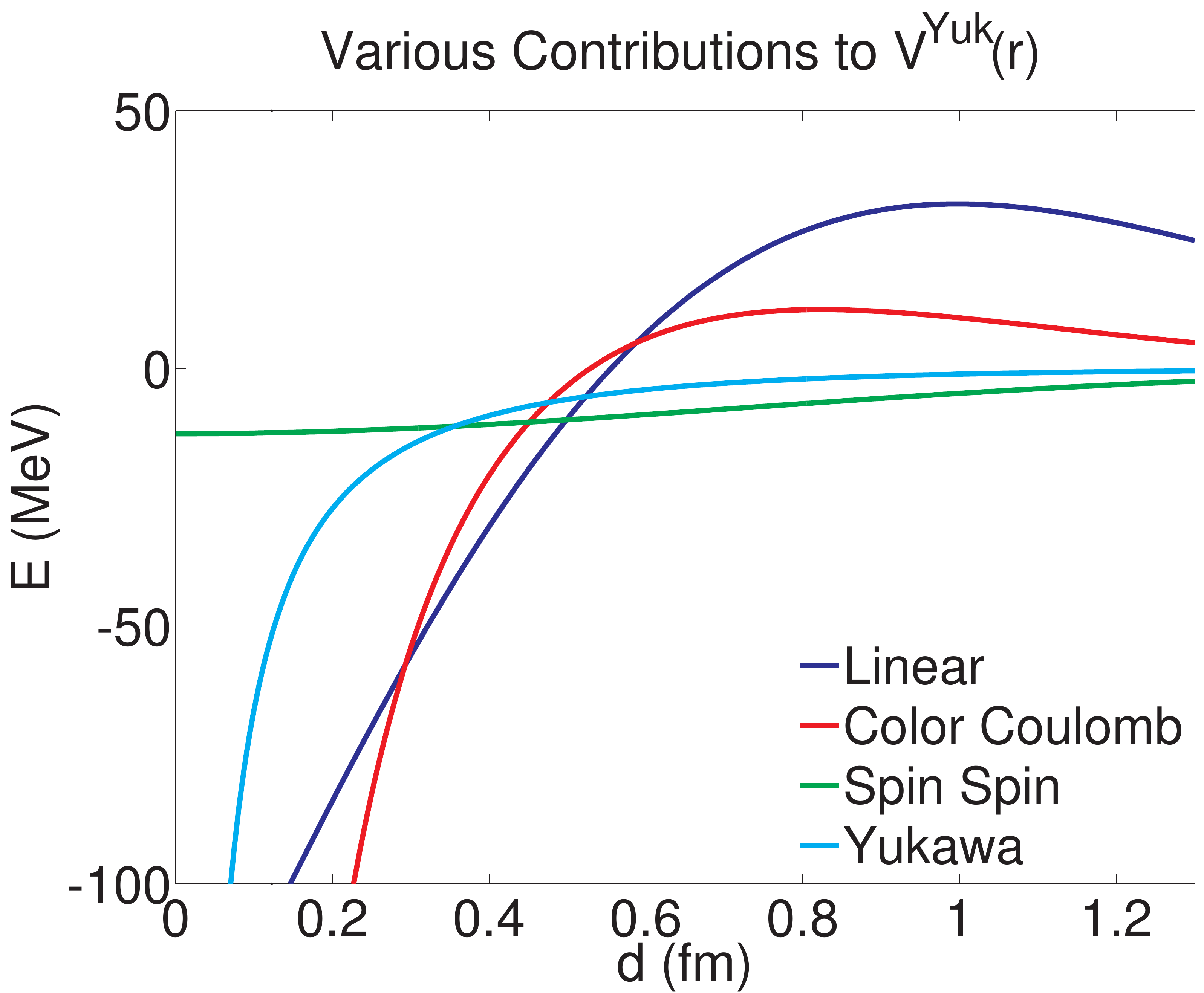}
\caption{Contribution of image charges to the potential (left) and contributions to the potential model $V_{HLHL}$ from the individual terms in eq.~\ref{eqn:pot_mod}}
\label{fig:Potential_Contributions}
\end{figure}

In principle, interactions with each of the infinitely many image charges contribute to the potential and must be included. In practice however, we may restrict ourselves to contributions where the image of the first meson is $\leq 3L/2$ ($\sim 4.5$ fm) away from the second and vice versa. This approximation is valid as the contribution of these truncated images (at separations of $r>3L/2$) to the potential (with the choice of parameters outlined below) 
is $\mathcal{O}\left(10^{-4}\right) \text{MeV}$. 
With the inclusion of the image charges our potential model then becomes:
\begin{align}
V_{Im}^{Yuk}= V^{Yuk}\left(r\right) + 2\sum_{r_{i} < 3L/2} V^{Yuk}\left(r_i\right)
\end{align}
The addition of these image charges modify the potential at long distance as illustrated in Fig. \ref{fig:Potential_Contributions}

\begin{figure}[]
\centering
\includegraphics[height=10cm]{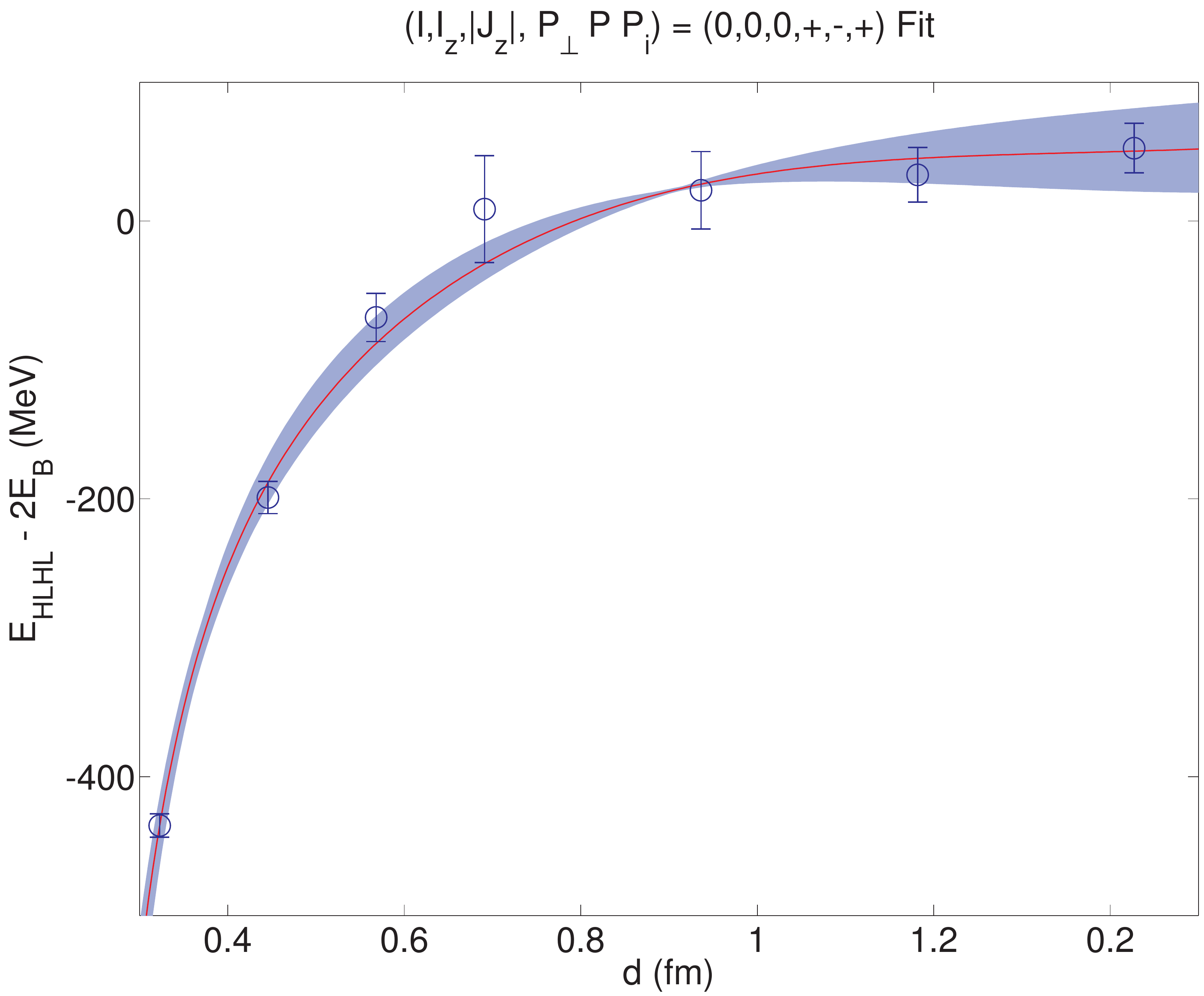}
\caption{Fit of the potential model in eq.~\ref{eq:ourModel} to the $\left(I,I_z,|J_z|,P_{\perp},P,P_i\right) = \left(0,0,0,+,-,+\right)$ channel. The colored band represents the uncertainty in the fit paramters $\beta$ and $g$ from jackknife analysis.}
\label{fig:Pot_0000_Fit}
\end{figure}

The final modification made to the potential model is a modification of the spin dependent coefficients $C_I$ and $C_{\bold{S} \cdot \bold{S}}$. The original presentation of this phenomenological potential model in Ref. \cite{Barnes} sought to provide a comparison with the lattice calculations of the time, which had an incomplete classification of the HLHL states in terms of the total isospin $I$ and spin $S$ of the system, 
while also maintaining a connection with the physical B meson states. Because of this, classification of the various potentials was made in terms of the physical $B$ and $B^*$ (first angular excitation of the $B$ meson) with respect to the quantum numbers $I$ and $S$. 

The difference in angular momentum spaces of the non-static and static limit prevents a direct interpretation of the lattice data from the present work in terms of physical $B$ and $B^*$ states, and our classification of states makes it difficult to reconcile the previous classification with ours. 
We therefore choose to recalculate the spin dependent coefficients of the potential model relevant for the static limit BB system we study on the lattice, the results of which are presented in Table \ref{tab:Fit_Params} (For details of the calculation, see Appendix \ref{App:coef}).
The previous determination of these coefficients for the HLHL system included spin degrees of freedom for the heavy quarks in the two meson states $| M_{i} M_{j}>$ allowing for better classification of the potential in terms of non-static limit states. We choose to neglect the spin degrees of freedom of the heavy quarks in our determination, effectively fully implementing the static limit for the the potential model. Thus the spin degrees of freedom of our two meson kets $| M_{i} M_{j} >$ are just those of the spin of the light degrees of freedom of our HLHL state. The evaluation of these coefficients however requires knowledge of the total angular momentum of the two meson state, a point that has been neglected until now. As we seek to fit the $\left(I,I_z,|J_z|,P_{\perp},P,P_i\right) = \left(0,0,0,+,-,+\right)$ and $\left(1,1,0,-,-,+\right)$ states, we need to determine if these particular states are in a symmetric angular momentum triplet, or an antisymmetric angular momentum singlet. In order to make this identification, we must rely on the overall symmetries of the state in question. We know that the parity $P$ of a given state is the product of the intrinsic parity $P_i$ and the symmetry of the spatial wavefunction. From this relationship, and with knowledge of the symmetry of the isospin spatial wavefunction, we can infer the symmetry of the angular momentum wavefunction:
\begin{align}
Sym_J = \left(-\right)\left(Sym_I\right) \left(P_i\right) \left(P\right),
\end{align} 
where $Sym_J$ and $Sym_I$ the symmetries of the angular momentum and isospin wavefunctions. The overall negative sign appears from exchanging fermions in the parity operation. Using this we are able to identify the $\left(I,I_z,|J_z|,P_{\perp},P,P_i\right) = \left(0,0,0,+,-,+\right)$ channel with $Sym_J = -$ as a $J=0$ state, and the $\left(I,I_z,|J_z|,P_{\perp},P,P_i\right) = \left(1,1,0,-,-,+\right)$ channel with $Sym_J = +$ as a $J=1$ state. The spin dependent coefficients can then be recalculated for our states and are shown in Table [\ref{tab:Fit_Params}].

\subsection{Fitting Procedure and Bound State Determination}
In fitting the potential model of eq.~\ref{eq:ourModel} to our lattice data, we use two free fit parameters: $\beta$ and $g$ and take the remaining parameters $b$, $\bar{m}$ and $\alpha_s$ to be $0.18 \text{GeV}^2$, $0.33 \text{GeV}$, and $0.5$ respectively as in Ref. \cite{Barnes}. 
A fit is performed for each of 305 jackknife ensembles, allowing for an accurate way to estimate the error on the extracted fit parameters, shown in Table [\ref{tab:Fit_Params}]. As we are ultimately interested in the energy levels allowed by the potential model, and not the model parameters themselves, we will only briefly comment on the fit parameters. It is immediately obvious that $g$ is not well determined for the $J=1$ channel. It's also interesting that the fit parameter $\beta$ is significantly smaller for the $J=0$ channel, indicating a much narrower spatial distribution of the two meson wavefunction.

Once the fit parameters have been extracted they are then inserted into the two body radial Schrodinger equation to determine if any bound states exist. As we are restricting ourselves to $L=0$ states, the two body Schrodinger equation can be written as:
\begin{align}
\left[-\frac{\hbar^2}{2 \bar{m}} \frac{d^2}{dr^2} + V^{Yuk}\left(r\right)\right]u\left(r\right) = E u\left(r\right)
\label{eqn:schrod}
\end{align}

where $\bar{m}$ is the reduced mass of a two B meson system (with the single meson mass taken from the Particle Data Group \cite{PDG}), $u\left(r\right) = r\Psi\left(r\right)$ and $V^{Yuk}\left(r\right)$ is the potential model presented in the preceding section excluding the image terms.  

Eq.~\ref{eqn:schrod} is then solved numerically as an eigenvalue problem with a spatial discretization of 0.01 fm and a spatial cutoff of 10 fm (corresponding to a sphere with $r=10$ fm), and the boundary condition that $\Psi\left(r\right)\bigg|_{r=10} = 0$. This spatial volume provides ample space for the potential to decay to zero. The eigenvalue spectrum is then analyzed for each of the two states discussed above. While the $J=1$ channel exhibits a near continuum of positive eigenvalues (discrete only because of the numerical solution method), the $J=0$ channel does admit a single bound state with energy $E_0 = -50.0(5.1)$ MeV (with the uncertainty determined by carrying through the jackknife analysis from the fit parameters and solving eq.~\ref{eqn:schrod} for each of the 305 $\left(\beta,g\right)$ sets). 
Aside from the binding energy, we can also calculate the RMS radius for the two meson wavefunction $\Psi\left(r\right)$ from the wavefunctions $u\left(r\right)$ above:

\begin{align}
r_{RMS} \equiv \left<r^2\right>^{1/2} =\left[\frac{\sum_i r_i^2 \left|u\left(r_i\right)\right|^2}{\sum_i\left|u\left(r_i\right)\right|^2}  \right]^{1/2}
\end{align}

For the bound state wavefunction $u_0\left(r\right)$, we find an RMS radius of 0.383(6) fm, the error again estimated by jackknife analysis.

Although no previous calculation of the binding energy in this particular static-limit channel exists (lattice or otherwise), Ref. \cite{Vijande2009} does quote binding energies and RMS radii for a doubly bottom $J^P \left(L, S, I\right) = 0^+ \left(0,0,0\right)$ channel which is consistent (in the static limit) with the quantum numbers of our static limit $\left(I,I_z,|J_z|,P_{\perp},P,P_i\right) = \left(0,0,0,+,-,+\right)$ channel. This reference uses two different potential models to calculate binding energies: the constituent quark cluster model CQC and the the Bhaduri-Cohler-Nogami or BCN model. 
The BCN model includes the same interactions as those used in Ref. \cite{Barnes} to derive the potential used to fit our lattice results (namely, color coulomb, linear confinement and spin-spin).
Furthermore, the BCN parameters corresponding to string tension $b$, strong coulpling $\alpha_s$, and constituent quark mass $\bar{m}$ used in \cite{Vijande2009} are very similar to those used in our potential model (compare our $\left(b, \alpha_s,\bar{m}\right) = \left(0.18\;\text{GeV}^2, 0.5, 0.33 \;\text{GeV}\right)$ to $\left(0.186 \;\text{GeV}^2, 0.52, 0.337 \;\text{GeV}\right)$).
These binding energies should provide a relevent point of comparison for our results provided our lattice discretization errors have minor effects on the extracted potential model fit parameters. In comparison, we find our values for the binding energy and RMS radius to be consistent with the values quoted in \cite{Vijande2009} from the BCN model 
$ \left(E_0, r_{RMS}\right) = \left(-52 \text{MeV}, 0.334 \text{fm}\right)$, providing a good cross check that our lattice calculation has identified a bound state in the static limit $\left(I,I_z,|J_z|,P_{\perp},P,P_i\right) = \left(0,0,0,+,-,+\right)$ channel. 
The fact that the bound state identified in that work has an RMS radius that is smaller than the sum of the individual mesonic RMS radii is indicative  of the compact nature of that bound state. 
Additionally, as illustrated in Ref. \cite{Wagner2010} (see eqns. 4), the static limit HLHL tetraquark state can be written as a linear combination of products of two single meson wavefunctions in different spin states. This is consistent with the idea that although the compact tetraquark state may have a complicated color space structure composed of color vectors, this state can always be decomposed into a linear combination of products of two single meson wavefunctions.

\begin{table}[h]
\centering
\begin{tabular}{|c||c|c|c|c|c|c|}
\hline
$\left(J, J_z\right)$ &$C_I$ & $C_{\bold{S} \cdot \bold{S}}$ & $\beta$ (GeV)& $g$ & $\chi^2/d.o.f.$ & $E_0$ (MeV)\\
\hline
\hline
(0,0) &  -1 & 3/4 & 0.274(14) & 0.041(12) & 0.9943 & -50.0(5.1)\\
\hline
(1,0) & 1 & 1/4 & 0.459(38) & 0.016(20) & 0.4119 & N/A\\
\hline
\end{tabular}
\caption{Spin dependent coefficients from reference \cite{BarnesCoef} and fit parameters from fitting our lattice data to a modified version of the model presented in ref. \cite{Barnes}. Here $\beta$ corresponds to the spatial width of the HL meson wavefunction, and $g$ is the coupling strength of the additional Yukawa term introduced in this work. The uncertainties quoted for the fit parameters are jackknife estimates.}
\label{tab:Fit_Params}
\end{table}

\section{Conclusions}
We have computed using lattice QCD the interaction potential between two b-meson states in the limit of static b quarks. With this lattice potential parametrized with a functional form
motivated by the quark model description of the two b-meson interaction, we have 
 determined the bound state energies in the heavy-light-heavy-light (HLHL) tetraquark system. 
 To perform this study we introduced colorwave propagators for calculating meson correlation functions and extended the formalism to the HLHL system in order to provide a novel way for an efficient calculation of HLHL correlation functions for several $\left( I,I_z,\left| J_z\right|, P_{\perp}, P, P_{i}\right) $ channels. The effect of limiting the colorwave plane wave basis on the ground state overlap of single HL correlation functions was explored,
 and a choice for the momentum cutoff  $p^2_{cut}$ was made to optimize the quality of the signal versus the computational cost. 
 For a single HL meson, results indicate that a more localized interpolating field has a better overlap on the ground state, suggesting the compact nature of the HL meson.
 
HLHL potentials were calculated for 24 distinct $\left( I,I_z,\left| J_z\right|,P_{\perp},P,P_{i}\right) $ channels, exhibiting three distinct asymptotic values as $r \rightarrow \infty$ corresponding to the different ways $B$ and $B_1$ mesons can be combined. 
The tendency of the HLHL energy to overshoot the expected asymptotic value of $E_{B_1} + E_B$ and $2E_{B_1}$ may be due to contamination from excited states and the possibility of $B_1$ mixing with a $B - \pi$ state.
It was determined that the attractiveness or repulsiveness of the HLHL potential corresponds directly to the symmetry of the two meson spatial wavefunction under spatial inversion, in agreement with Ref. \cite{Wagner2011}. The asymptotic behavior of the various HLHL states was shown to be dependent on the intrinsic parity of the state. While the $P_i = -$ states have only one asymptotic value (corresponding to a single two meson $B B_1$ component), the $P_i = +$ channels have two asymptotic values corresponding to both $B B$ and $B_1 B_1$ two meson components. By examining the construction of single HL correlation functions, it was determined that we could increase overlap with the $B B$ and $B_1 B_1$ two meson wavefunctions by projecting the correlation functions to include only positive or negative parity components of the Dirac basis quark spinors. 

The existence of bound states was then explored for the $\left(I,I_z,|J_z|,P_{\perp},P,P_i\right) = \left(0,0,0,+,-,+\right)$ channel as it exhibited a wider and deeper potential when compared with the other attractive potentials. Analysis was also carried out for the $\left(I,I_z,|J_z|,P_{\perp},P,P_i\right) = \left(1,1,0,-,-,+\right)$ for the purposes of comparison. A modified version of the potential model described in Ref. \cite{Barnes} was used to fit the lattice data, and two fit parameters $\beta$ (the gaussian width of the HL meson wavefunction) and $g$ (the Yukawa interaction constant) were extracted from,
 the fit. Inserting the potential  with the  extracted fit parameters into the two body Schr\"odinger equation, we then solved numerically for the eigenvalues of the hamiltonian, searching for any negative energy eigenstates. A single negative energy bound state was found in the $\left(0,0,0,+,-,+\right)$ channel, with an energy of $E_0 = -50.0(5.1)$ MeV and RMS radius $r_{RMS} = 0.383(6)$ fm. These results were found to be consistent with results presented in Ref. \cite{Vijande2009} for the state $J^P \left(L, S, I\right) = 0^+ \left(0,0,0\right)$ (which maps onto our $\left(0,0,0,+,-,+\right)$ channel in the static limit). 
The errors quoted on these results are statistical only. One needs to account for several systematic errors such as $1/m_b$ corrections ($m_b$ the b quark mass), lattice spacing effects as well as dependence on the light quark mass.~\footnote{Just before  completion of this manuscript a study of the
same system appeared as a preprint~\cite{Bicudo:2012qt}. Some of these systematics were studied there.}


\acknowledgments
We would like to thank W. Detmold, S. Meinel, and E. Mastropas for useful discussions. This work was supported in part by DOE grants DE-AC05-06OR23177 (JSA) and DE-FG02-04ER41302 and  DE-FG02-04ER41302 as well as the JSA Jefferson Lab Graduate Fellowship Program. 

\appendix

\section{Parity content of HL interpolating operators}
\label{App:Single_HL}
Here we show that correlation functions for our $B$($B_1$) states are composed entirely of upper(lower) components of the Dirac basis components of the light quark flavors. We begin with a general HL correlation function with arbitrary source and sink operators (neglecting color indices and working in the Dirac basis): 

\begin{align}
C_{HL}\left(t\right)_{i,j} &= \sum_{\vec{x}} \left< \mathcal{O}_{B_i}\left(\vec{x},t\right) \mathcal{O}_{B_j}^{\dagger}\left(\vec{x},0\right)\right> \nonumber\\ 
				   &= \sum_{\vec{x}} \left< \bar{Q}\left(\vec{x},t\right) \Gamma_i q\left(\vec{x},t\right) \bar{q}\left(\vec{x},0\right) \Gamma_j Q\left(\vec{x},0\right) \right> \nonumber\\
				    &= \sum_{\vec{x}} tr\left( \gamma_5 \left(S_H\left(\vec{x},t;0\right)\right)^{\dagger} \gamma_5 \Gamma_{i} S_L\left(\vec{x},t;0\right)\Gamma_{j}\right)
\label{eqn:HL_Corr}
\end{align}

$S_{H}$ is a heavy quark propagator given by:
\begin{equation}
S_H\left(\vec{x},t;t_0\right) = \left( \frac{1+\gamma_4}{2}\right) W\left(\vec{x},t;t_0\right) = P_+ W\left(\vec{x},t;t_0\right)
\end{equation} 
where $ W\left(\vec{x},t;t_0\right) $ is a Wilson line from $t_0$ to $t$. Substituting this, we have:
\begin{align}
C_{HL}\left(t\right) &= \sum_{\vec{x}} tr\left( \gamma_5 \left(P_+ W^\dagger\left(\vec{x},t;0\right)\right) \gamma_5 \Gamma_{i} S_L\left(\vec{x},t;0\right)\Gamma_{j}\right) \nonumber \\
				   &= \sum_{\vec{x}} tr_c\left( W^\dagger\left(\vec{x},t;0\right) tr_d \left(\Gamma_{i} P_- \Gamma_{j} S_L\left(\vec{x},t;0\right)\right)\right)
\end{align}
 where we have used $\gamma_5 P_+ \gamma_5 = P_-$. For $\Gamma_{i} = \Gamma_{j} = 1$, we project to only the lower components of the Dirac basis light quark propagator, while for $\Gamma_{i} = \Gamma_{j} = \gamma_5$ we project only to the upper components of the Dirac propagator. 

\section{Construction of light quark wavefunctions}
\label{App:wfn}
To determine two quark wavefunctions in spin and flavor space yielding the quantum numbers $\left(I,I_z,|J_z|, P_{\perp},  P,  P_i\right)$, we begin with states of definite $\left(I,I_z,J,J_z,P_i\right)$:  
\begin{align}
\left[q_1\left(p_1\right)  q_2\left(p_2\right)\right]\bigg|_{\left( I,I_z,J,J_z,P_i\right)}   = \sum_{\substack{m_1,m_2 \\ t_1,t_2}} W_{m_1,m_2}^{J,J_z} W_{t_1,t_2}^{I,I_z} q_1\left(m_1,t_1,p_1\right) q_2\left(m_2,t_2,p_2\right)
\end{align}
where $m,t,p$ are the projections of spin and isospin along the z-axis and the intrinsic parities of the light quarks, and the $W_{m_1,m_2}^{J,J_z} = \left<1/2,m_1,1/2,m_2|J,J_z\right>$, $W_{t_1,t_2}^{I,I_z} = \left<1/2,t_1,1/2,t_2|I,I_z\right>$ are the Clebsch-Gordon for angular momentum and isospin. From these operators, we average over $J_z=\pm1$ states and determine $P_{\perp}$ from the quantum numbers $P_i$ and $P$ and the spatial symmetry of the operator.
It should be noted here that there are two combinations of $\left(p_1,p_2\right)$ that contribute to the $P_i = +1$ HLHL states, and we make the decision to keep these as distinct operators.

Linear combinations of the above operators are then taken to produce states of definite exchange parity $P$, the necessary combinations determined by summing over sets of the above operators that map onto each other under $P$ with the appropriate weight $W_{p_1,p_2}^P=\pm 1$ 
\begin{align}
\left[q_1 q_2\right]\bigg|_{\left(I,I_z,|J_z|, P_{\perp},  P,  P_i\right)} = \sum_{p_1,p_2} W_{p_1,p_2}^P \left[q_1\left(p_1\right)  q_2\left(p_2\right)\right]\bigg|_{\left( I,I_z,|J_z|, P_{\perp}, P_i\right)}
\end{align}

\section{Determination of spin coefficients for potential model}
\label{App:coef}
\begin{figure}[]
\centering
\includegraphics[height=5cm]{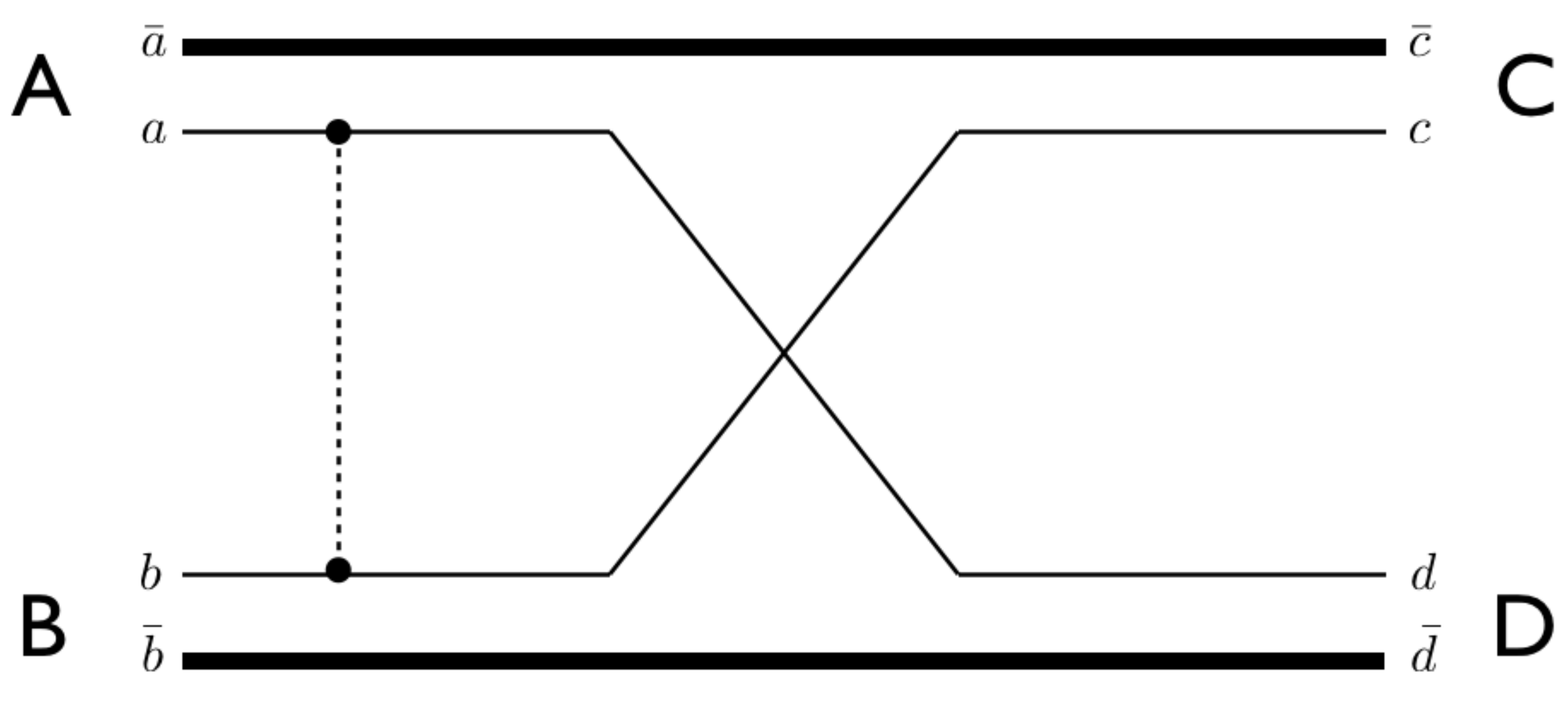}
\caption{Transfer 2 diagram from Ref. \cite{Barnes}}
\label{fig:Coef_Diagram}
\end{figure}

Here we present our derivation of the spin coefficients $C_I$ and $C_{\bold{S} \cdot \bold{S}}$ presented in Table \ref{tab:Fit_Params}. In Ref. \cite{BarnesCoef}, an interaction potential for two meson states is calculated by including spin-spin, color coulomb and linear confinment interactions in a two quark interaction hamiltonian. 
By considering these interactions between each of the quark quark pairs in a 4 quark (2 meson) scattering state, transfer matrix elements are calculated and then Fourier transformed to give a corresponding position space potential. In Ref. \cite{Barnes}, this method was applied to the HLHL system. 
When calculating the spin dependent portion of the potential, all but one of the interaction diagrams (referred to as ``Transfer 2'') can be neglected because the spin of the heavy quarks is neglected in the static approximation. This diagram includes an insertion of the interaction hamiltonian between the two light quarks, as illustrated in in Fig. \ref{fig:Coef_Diagram}. The spin dependent contribution of this diagram to the potential can be factorized such that all the dependence enters through two coefficients, which are defined as matrix elements between the initail and final two meson states:

\begin{align}
C_I &= \left< C D | I | A B \right>  \\
C_{\bold{S} \cdot \bold{S}} &= \left< C_{i} D_{j} | \bold{S}_i \cdot \bold{S}_j | A_{i} B_{j} \right>
\end{align}
Where $I$ here is understood to be the identity operator in spinor space. Upon inspection of the diagram, it's clear that the matrix element of $I$ will not always trivially be unity due to the quark interchange between the initial and final two meson state. 

With respect to Fig.~\ref{fig:Coef_Diagram}, these matrix elements as outlined in \cite{BarnesCoef} are defined explicitly as:

\begin{align}
C_I &= \left< C D | I | A B \right> \nonumber \\
    &=  \chi_{s_c,s_{\bar{c}}}^{\lambda_C}\chi_{s_d,s_{\bar{d}}}^{\lambda_D}
        \left[ <s_c, s_d|\bold{I} | s_a, s_b > \delta_{s_{\bar{a}},s_{\bar{c}}} \delta_{s_{\bar{b}},s_{\bar{d}}}\right]
        \chi_{s_a,s_{\bar{a}}}^{\lambda_A}\chi_{s_b,s_{\bar{b}}}^{\lambda_B}\\
C_{\bold{S} \cdot \bold{S}} &=  \left< C_i D_j | \bold{S}_i \cdot \bold{S}_j | A_i B_j \right> \nonumber \\
              &=\chi_{s_c,s_{\bar{c}}}^{\lambda_C}\chi_{s_d,s_{\bar{d}}}^{\lambda_D}
\left[ <s_c, s_d| \bold{S}_i \cdot \bold{S}_j| s_a, s_b > \delta_{s_{\bar{a}},s_{\bar{c}}} \delta_{s_{\bar{b}},s_{\bar{d}}}\right]
\chi_{s_a,s_{\bar{a}}}^{\lambda_A}\chi_{s_b,s_{\bar{b}}}^{\lambda_B}
\end{align}

For our purposes, we wish to entirely neglect the spin of the heavy quarks in the above matrix elements. 
Because of this, the Clebsch-Gordan coefficients  $\chi_{s_c,s_{\bar{c}}}^{\lambda_C}$ etc. (relating the spin of the two quark state to the meson state) are all unity.
The states between which we wish to calculate these matrix elements are two particle angular momentum eigenstates $\left|J, J_z\right>_{a,b} \equiv \left|s_a, s_b\right>\bigg|_{J,J_z}$, of which we are only interested in $\left|1,0\right>$ and $\left|0,0\right>$. 
To account the light quark exchange in Fig. \ref{fig:Coef_Diagram}, we note the following relations:
\begin{align}
\left|s_a, s_b\right>\bigg|_{\substack{J=0, \\ J_z=0}} = \frac{1}{\sqrt{2}}\left( \left|\uparrow \downarrow\right> - \left|\downarrow \uparrow\right>\right) =  -\frac{1}{\sqrt{2}}\left( \left|\downarrow \uparrow\right> - \left|\uparrow \downarrow\right>\right) =  -\left|s_c, s_d\right>\bigg|_{\substack{J=0, \\ J_z=0}}
\end{align} 
and 
\begin{align}
\left|s_a, s_b\right>\bigg|_{\substack{J=1, \\ J_z=0}} = \frac{1}{\sqrt{2}}\left( \left|\uparrow \downarrow\right> + \left|\downarrow \uparrow\right>\right) = \frac{1}{\sqrt{2}}\left( \left|\downarrow \uparrow\right> + \left|\uparrow \downarrow\right>\right) =\left|s_c, s_d\right>\bigg|_{\substack{J=1, \\ J_z=0}}
\end{align} 

From the above relations, it is easy to calculate the matrix elements of interest for our problem (for the states $\left|1,0\right> \rightarrow \left|1,0\right>$ and $\left|0,0\right> \rightarrow \left|0,0\right>$):
\begin{align}
\left<1,0\right|_{c,d} \bold{I} \left|1,0\right>_{a,b} &= \left<1,0\right|_{a,b} \bold{I} \left|1,0\right>_{a,b} = 1 \\ 
\left<0,0\right|_{c,d} \bold{I} \left|0,0\right>_{a,b} &= (-)\left<0,0\right|_{a,b} \bold{I} \left|0,0\right>_{a,b} = -1
\end{align}
and 
\begin{align}
\left<1,0\right|_{c,d}  \bold{S}_i \cdot \bold{S}_j \left|1,0\right>_{a,b} &= \left<1,0\right|_{a,b}  \bold{S}_i \cdot \bold{S}_j \left|1,0\right>_{a,b} = 1/4 \\ 
\left<0,0\right|_{c,d}  \bold{S}_i \cdot \bold{S}_j \left|0,0\right>_{a,b} &= -\left<0,0\right|_{a,b} \bold{S}_i \cdot \bold{S}_j \left|0,0\right>_{a,b} = -(-3/4)
\end{align}

\bibliographystyle{ieeetr}
\bibliography{HLHLBIB}

\end{document}